\documentclass[12pt,a4paper]{article}
\usepackage[utf8]{inputenc}
\usepackage{epsf}
\usepackage{latexsym,amssymb,euscript}
\usepackage[dvips]{graphicx}
\usepackage[numbers,sort&compress]{natbib}
\usepackage{amsmath}
\usepackage{nicefrac}
\usepackage{slashed}
\usepackage{booktabs}
\usepackage{hyperref}
\usepackage{braket}
\usepackage{chngcntr}
\usepackage{bbold}
\usepackage{multicol}
\usepackage{graphics}
\usepackage{graphicx}
\usepackage{mciteplus}
\usepackage{caption}
\usepackage{subcaption}
\usepackage{pdfpages}
\usepackage[titletoc]{appendix}
\graphicspath{{./figures/}}
\hypersetup{
 linktocpage = true,
 urlcolor = urlblue,
 colorlinks = true,
 linkcolor = urlblue,
 anchorcolor = urlblue,
 citecolor = urlblue,
 pdfstartview = {XYZ null null 1.25} 
           }
\usepackage[left=2cm, right=2cm]{geometry}
\usepackage{pstricks}
\usepackage{color}
\usepackage{xcolor}
\definecolor{urlblue}{rgb}{0.2,0.4,0.7}
\definecolor{citegreen}{rgb}{0,0.4,0.2}
\definecolor{linkred}{rgb}{0.9,0.2,0.1}
\usepackage{float}
\usepackage{academicons}
\definecolor{orcidlogocol}{HTML}{A6CE39}
\usepackage{fancyhdr}
\newcommand{\drv}{{\rm d}}

\newcommand{\LQCD}{\Lambda_{\rm QCD}}
\newcommand{\MSb}{\overline{\rm MS}}

\newcommand{\LL}{{\rm LL/LO}}
\newcommand{\NLL}{{\rm NLL/NLO}}
\newcommand{\NLLp}{{\rm NLL/NLO^+}}
\newcommand{\HENLOp}{{\rm HE}\mbox{-}{\rm NLO^+}}
\newcommand{\HENLO}{{\rm HE}\mbox{-}{\rm NLO}}
\newcommand{\CmLL}{{\cal C}_m^\LL}

\newcommand{\CmNLLp}{{\cal C}_m^\NLLp}

\newcommand{\CmHENLOp}{{\cal C}_m^{{\rm HE}\text{-}{\rm NLO}^+}}
\newcommand{\DY}{\Delta Y}

\newcommand{\tcite}[1]{~\cite{#1}}
\newcommand{\tref}[1]{~\ref{#1}}
\newcommand{\eref}[1]{~\eqref{#1}}

\newcommand{\tarr}{
\begin{array}}
\newcommand{\earr}{\end{array}}

\newcommand{\orcidFGC}{\href{https://orcid.org/0000-0003-3299-2203}{\includegraphics[scale=0.1]{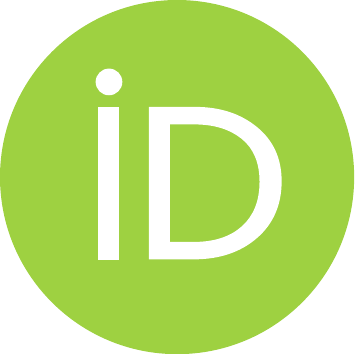}}}

\begin{document}

\begin{titlepage}

\begin{center}
  {\LARGE \bf Emergence of high-energy dynamics \vskip.025cm from cascade-baryon detections at the LHC}
\end{center}

\vskip 0.35cm

\centerline{
Francesco Giovanni~Celiberto$^{\;1,2,3,4\;\ddagger}$ \orcidFGC
}

\vskip .4cm

\centerline{${}^1$ {\sl European Centre for Theoretical Studies in Nuclear Physics and Related Areas (ECT*),}}
\centerline{\sl I-38123 Villazzano, Trento, Italy}
\vskip .18cm
\centerline{${}^2$ {\sl Fondazione Bruno Kessler (FBK), %}}
%\centerline{\sl
I-38123 Povo, Trento, Italy} }
\vskip .18cm
\centerline{${}^3$ {\sl INFN-TIFPA Trento Institute of Fundamental Physics and Applications,}}
\centerline{\sl I-38123 Povo, Trento, Italy}
\vskip .18cm
\centerline{${}^4$ {\sl Universidad de Alcal\'a (UAH), E-28805 Alcal\'a de Henares, Madrid, Spain} }
\vskip 0.55cm

\begin{abstract}
\vspace{0.25cm}
\hrule \vspace{0.50cm}
We propose the inclusive detection at the LHC of a cascade $\Xi^-/\bar\Xi^+$ baryon in association with a jet, as a novel probe channel for the QCD dynamics at high energies.
We investigate the behavior of a selection of distributions, differential in rapidity, azimuthal angle and/or transverse momenta, calculated via the hybrid high-energy/collinear factorization encoding the full next-to-leading BFKL resummation of energy logarithms. We come out with the conclusion that the fragmentation mechanism underlying the production of $\Xi^-/\bar\Xi^+$ baryon states leads to stabilization effects of the resummation, similar to those recently observed in the context of heavy-flavor studies within the same formalism.
\vspace{0.25cm} \hrule
\vspace{0.75cm}
{
 \setlength{\parindent}{0pt}
 \textsc{Keywords}: \vspace{0.15cm} \\ High-energy QCD \\ Resummation \\ Natural stability \\ $\Xi$ baryons \\ Cascade particles
%\vspace{0.15cm} %\hrule
%\vspace{0.50cm}
}
\end{abstract}

%\vskip .5cm
\vfill
%\hrule
$^{\ddagger}${\it e-mail}:
\href{mailto:francesco.celiberto@uah.es}{francesco.celiberto@uah.es}

%$^{\dagger\dagger}${\it e-mail}:
%\href{mailto:d-ivanov@math.nsc.ru}{d-ivanov@math.nsc.ru}

%$^{\S}${\it e-mail}:
%\href{mailto:mohammed.maher@unical.it }{mohammed.maher@unical.it }

%$^{\P}${\it e-mail}:
%\href{alessandro.papa@fis.unical.it}{alessandro.papa@fis.unical.it}

\end{titlepage}

%-----------------------------------------
\section{Opening remarks}
\label{sec:intro}
%-----------------------------------------

The study of the dynamics of Quantum ChromoDynamics (QCD) in the high-energy domain is a core research field at the Large Hadron Collider (LHC) as well as of new-generation accelerators and facilities\tcite{Chapon:2020heu,Anchordoqui:2021ghd,Feng:2022inv,Celiberto:2022rfj,Hentschinski:2022xnd,Accardi:2012qut,AbdulKhalek:2021gbh,Khalek:2022bzd,Acosta:2022ejc,AlexanderAryshev:2022pkx,Brunner:2022usy,Arbuzov:2020cqg,Abazov:2021hku,Bernardi:2022hny,Amoroso:2022eow,Celiberto:2018hdy,Klein:2020nvu,2064676,MuonCollider:2022xlm,Aime:2022flm,MuonCollider:2022ded,Accettura:2023ked,Begel:2022kwp}. 
In the Regge--Gribov or \emph{semi-hard} regime\tcite{Gribov:1983ivg,Celiberto:2017ius}, namely where a stringent scale hierarchy, $\sqrt{s} \gg \{Q\} \gg \LQCD$, ($s$ is the center-of-mass energy squared, $\{Q\}$ represents one or a set of hard scales typical of the process, and $\LQCD$ is the QCD hadronization scale) is stringently preserved, large $\ln(s/Q^2)$ type logarithms become relevant. They enter the perturbative expansion with a power growing with the order of the strong coupling, $\alpha_{s}$.
The convergence of the perturbative series needs to be restored by accounting for those large energy logarithms.
The most adequate formalism to perform such an all-order resummation is the Balitsky--Fadin--Kuraev--Lipatov (BFKL) approach~\cite{Fadin:1975cb,Kuraev:1976ge,Kuraev:1977fs,Balitsky:1978ic},  
allows us to catch all terms proportional to $(\alpha_s\ln(s))^n$, in the leading-logarithmic (LL) approximation, and of those of the form $\alpha_s^{n+1}\ln(s)^n$, in the next-to-leading logarithmic (NLL) approximation.

The BFKL-resummed cross section is built as a high-energy convolution between a Green's function, which determines the resummation of energy logarithms and it is independent from the considered final state, and two impact factors, describing the fragmentation of each incoming particles. The BFKL Green's function evolves according to an integral equation, whose kernel was computed within the next-to-leading order (NLO) for any fixed, not growing with $s$, momentum transfer $t$ and for any possible two-gluon exchange of color in the $t$-channel\tcite{Fadin:1998py,Ciafaloni:1998gs,Fadin:1998jv,Fadin:2000kx,Fadin:2000hu,Fadin:2000yp,Fadin:2004zq,Fadin:2005zj,Fadin:2023roz}.
Impact factors are instead process-dependent. Therefore, they are the most challenging building blocks of the cross section.
Only a few of them are known at the NLO: (\emph{a}) quarks and gluons\tcite{Fadin:1999de,Fadin:1999df,Ciafaloni:1998kx,Ciafaloni:1998hu,Ciafaloni:2000sq}, \emph{i.e.} the common basis to calculate (\emph{b}) forward-jet\tcite{Bartels:2001ge,Bartels:2002yj,Caporale:2011cc,Ivanov:2012ms,Colferai:2015zfa} and (\emph{c}) forward light-hadron\tcite{Ivanov:2012iv} impact factors, (\emph{d}) the impact factor for the light vector-meson leptoproduction, (\emph{e}) the ($\gamma^* \to \gamma^*$) impact factor\tcite{Bartels:2000gt,Bartels:2001mv,Bartels:2002uz,Bartels:2003zi,Bartels:2004bi,Fadin:2001ap,Balitsky:2012bs}, and (\emph{f}) the forward-Higgs impact factor in the large top-mass limit\tcite{Hentschinski:2020tbi,Celiberto:2022fgx}.

Remarkably, the high-energy resummation provided by BFKL gives us a chance to unveil the proton structure at low $x$ by means of single-forward production reactions.
The BFKL \emph{unintegrated gluon distribution} (UGD) in the proton reads as a convolution in the transverse-momentum space\tcite{Catani:1990xk,Catani:1990eg,Catani:1993ww,Ball:2007ra,Caola:2010kv} between the BFKL Green's function and a soft, nonperturbative quantity, known as proton impact factor.
First studies on the UGD were performed via deep-inelastic-scattering structure functions\tcite{Hentschinski:2012kr,Hentschinski:2013id} and light vector-meson helicity-dependent observables at HERA\tcite{Anikin:2009bf,Anikin:2011sa,Besse:2013muy,Bolognino:2018rhb,Bolognino:2018mlw,Bolognino:2019bko,Bolognino:2019pba,Celiberto:2019slj} and, more recently, at the Electron-Ion Collider (EIC)\tcite{Bolognino:2021niq,Bolognino:2021gjm,Bolognino:2022uty,Celiberto:2022fam,Bolognino:2022ndh}.
Then, the UGD was investigated through forward Drell--Yan\tcite{Motyka:2014lya,Brzeminski:2016lwh,Motyka:2016lta,Celiberto:2018muu} and single-forward quarkonium\tcite{Bautista:2016xnp,Garcia:2019tne,Hentschinski:2020yfm,Goncalves:2018blz,Cepila:2017nef,Guzey:2020ntc,Jenkovszky:2021sis,Flore:2020jau,ColpaniSerri:2021bla} emissions.
Taking advantage of the information on the gluon motion inside the proton encoded in the UGD, pioneering determinations of low-$x$ enhanced collinear parton distribution functions (PDFs) as well as of transverse-momentum-dependent (TMD) spin-dependent gluon TMDs were respectively obtained in Refs.\tcite{Ball:2017otu,Bonvini:2019wxf} and\tcite{Bacchetta:2020vty,Celiberto:2021zww,Celiberto:2022omz,Bacchetta:2021oht,Bacchetta:2021lvw,Bacchetta:2021twk,Bacchetta:2022esb,Bacchetta:2022crh,Bacchetta:2022nyv}.

First studies of high-energy observables accessible at the LHC in the spirit of BFKL where done in the context of the Mueller--Navelet emission of two jets produced at large transverse momenta and with a high distance in rapidity\tcite{Mueller:1986ey}. Here, a \emph{hybrid} high-energy/collinear factorization\tcite{Colferai:2010wu} was build to embody collinear inputs in the standard BFKL description (see Refs.\tcite{Deak:2009xt,vanHameren:2015uia,Deak:2018obv,VanHaevermaet:2020rro,Blanco:2020akb,vanHameren:2020rqt,Guiot:2021vnp,vanHameren:2022mtk} for another formalism, similar to our one).
Since Mueller--Navelet jet detections probe incoming protons at moderate values of longitudinal-momentum fractions, a PDF-based description is valid.
On the other side, however, high rapidity intervals lead to large transverse-momentum exchanges in the $t$-channel, so that energy logarithms are heightened.
Therefore, a hybrid factorization was proposed, where BFKL-resummed partonic hard factors are genuinely convoluted with collinear PDFs.
Several phenomenological analyses of Mueller--Navelet azimuthal-angle correlations were proposed so far\tcite{Caporale:2012ih,Ducloue:2013hia,Ducloue:2013bva,Caporale:2013uva,Caporale:2014gpa,Ducloue:2015jba,Celiberto:2015yba,Celiberto:2015mpa,Caporale:2015uva,Mueller:2015ael,Celiberto:2016ygs,Celiberto:2016vva,Caporale:2018qnm,deLeon:2021ecb,Celiberto:2022gji} 
and compared with the only experimental data available, \emph{i.e.} the CMS ones at $\sqrt{s} = 7\mbox{ TeV}$ and for \emph{symmetric} configurations of the transverse momenta of the two jets\tcite{Khachatryan:2016udy}.
Further studies of high-energy QCD via the hybrid factorization include: the inclusive detection of two light hadrons well separated in rapidity\tcite{Celiberto:2016hae,Celiberto:2016zgb,Celiberto:2017ptm,Celiberto:2017uae,Celiberto:2017ydk}, multi-jet emissions\tcite{Caporale:2015vya,Caporale:2015int,Caporale:2016soq,Caporale:2016vxt,Caporale:2016xku,Celiberto:2016vhn,Caporale:2016djm,Caporale:2016pqe,Chachamis:2016qct,Chachamis:2016lyi,Caporale:2016lnh,Caporale:2016zkc,Chachamis:2017vfa,Caporale:2017jqj}, hadron plus jet correlations\tcite{Bolognino:2018oth,Bolognino:2019cac,Bolognino:2019yqj,Celiberto:2020wpk,Celiberto:2020rxb}, Higgs plus jet rapidity and transverse-momentum distributions\tcite{Celiberto:2020tmb,Celiberto:2021fjf,Celiberto:2021tky,Celiberto:2021txb,Celiberto:2021xpm}, Drell--Yan plus jet tags\tcite{Golec-Biernat:2018kem}, heavy-flavored hadrons' hadroproductions\tcite{Boussarie:2017oae,Celiberto:2017nyx,Bolognino:2019ouc,Bolognino:2019yls,Bolognino:2019ccd,Celiberto:2021dzy,Celiberto:2021fdp,Bolognino:2021zco,Bolognino:2022wgl,Celiberto:2022dyf,Celiberto:2022grc,Bolognino:2022paj,Celiberto:2022keu,Celiberto:2022zdg,Celiberto:2022kza}, and heavy-light two jet systems\tcite{Bolognino:2021mrc,Bolognino:2021hxxaux}.
Among them, a study on $\Lambda$-baryons emissions possibly accompanied by light-jet detections provided us with an evidence that the tag of $\Lambda$ hyperons eases the comparison between theoretical results and experimental data for semi-hard observables\tcite{Celiberto:2020rxb}.
This is due to the lower statistics featured by the production of these baryons, which quenches the contamination of the so-called minimum-bias events.

One of the most relevant issues rising from the analysis of Mueller--Navelet final states is the weight of NLL corrections, which are of the same order, but generally with and opposite sign with respect to pure LL contributions. This brings to instabilities of the high-energy series that become strongly manifest when studies on renormalization and factorization scale variations are made.
As a result, differential cross distributions can easily become negative as of the rapidity separation between the two jets increases. Moreover, the high-energy description of observables sensitive to azimuthal-angle correlations turns out to be unphysical both in the small and the large rapidity range.
To cure these instabilities, several strategies have been proposed so far.
Among them, the Brodsky--Lepage--Mackenzie (BLM) procedure \cite{Brodsky:1996sg,Brodsky:1997sd,Brodsky:1998kn,Brodsky:2002ka}, as specifically designed for semi-hard reactions\tcite{Caporale:2015uva}, became very popular, since it allowed us to moderately suppress these instabilities on azimuthal-angle correlations and to slightly raise the agreement with experimental data. Unfortunately, employing BLM is fruitless on cross sections for light di-hadron and light hadron-jet observables. In particular, the found optimal renormalization scales are significantly larger than the natural ones suggested by kinematics\tcite{Celiberto:2017ius,Bolognino:2018oth,Celiberto:2020wpk}. This leads to a sizable loss of statistics for total cross sections. Therefore, any attempt at reaching the precision level was ineffective.

First, clear signals of a reached stability of the high-energy resummations under higher-order corrections and energy-scale variation were discovered only recently in LHC final states characterized by the production of particles with a large transverse mass, such as Higgs bosons~\cite{Celiberto:2020tmb,Celiberto:2021fjf,Celiberto:2021tky,Celiberto:2021txb}.
A striking result at the NLL level was achieved by studying again semi-hard observables sensitive to baryon detections, this time $\Lambda_c$ hadrons.
Strong stabilizing effects emerged in a study on double $\Lambda_c$ and $\Lambda_c$ plus jet emissions at the LHC\tcite{Celiberto:2021dzy}, and then on analogous observables sensitive to single-bottomed hadrons\tcite{Celiberto:2021fdp}.
Here, we provided a corroborating evidence that the peculiar behavior of variable-flavor-number-scheme (VFNS)\tcite{Mele:1990cw,Cacciari:1993mq} collinear fragmentation functions (FFs) depicting the production of those heavy-flavored hadrons at large transverse momenta\tcite{Kneesch:2007ey,Kniehl:2008zza,Kramer:2018vde,Kramer:2018rgb,Kniehl:2020szu,Anderle:2017cgl,Soleymaninia:2017xhc}
leads to a \emph{natural stabilization} of the high-energy series, with a substantial suppression of instabilities associated to higher-order corrections. The same stabilization pattern was then discovered also in the context of vector-quarkonium\tcite{Celiberto:2022dyf} and charmed $B$-meson\tcite{Celiberto:2022keu} final states studied by combining the BFKL resummation with collinear PDFs and nonrelativistic-QCD FFs \cite{Braaten:1993mp,Braaten:1993rw,Zheng:2019dfk,Zheng:2019gnb,Zheng:2021sdo}. This corroborated the statement that the natural stability is an \emph{intrinsic} feature shared by heavy-flavor emissions.

\begin{figure*}[!t]
\centering
\includegraphics[width=0.65\textwidth]{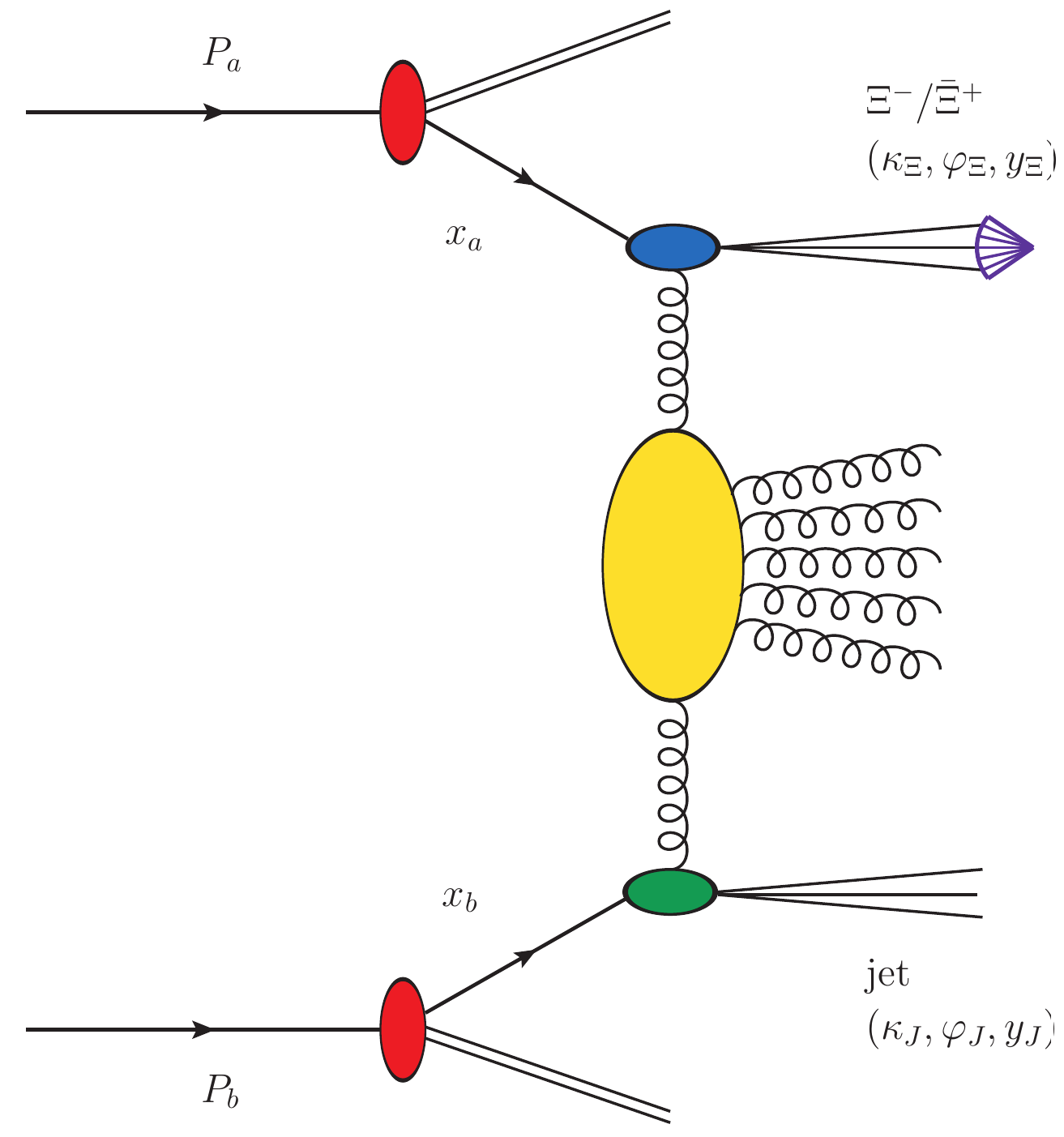}

\caption{Hybrid high-energy/collinear factorization for the inclusive $\Xi^-/\bar\Xi^+$~plus~jet detection. The blue (green) blob denotes the off-shell hard factor encoded in the hadron (jet) impact factor, whereas the indigo arrow depicts a $\Xi$ baryon emission via the fragmentation mechanism. The BFKL ladder, portrayed by the yellow blob, is connected to impact factors through Reggeon lines. The diagram was created with {\tt JaxoDraw 2.0}~\cite{Binosi:2008ig}.}
\label{fig:process}
\end{figure*}

In this article we consider the inclusive semi-hard detection of a $\Xi^-$ baryon, or its antiparticle $\bar\Xi^+$, accompanied by a jet, as a novel probe channel for the high-energy dynamics of QCD (see Fig.\tref{fig:process}). This reaction extends our program on baryon emissions at high energies, started with $\Lambda$ hyperons\tcite{Celiberto:2020rxb} and carried on with $\Lambda_c$ hadrons\tcite{Celiberto:2021dzy}.
The family of $\Xi$ particles consists of baryons whose lowest Fock state contains one up or one down quark and two other, more massive quarks.
Due to their highly unstable nature, they are also known as \emph{cascade} particles. Indeed they are typically observed to rapidly decay into lighter hadrons via a fast chain of decays, called cascade.
The existence of a neutral cascade hyperon, $\Xi^0$, as well as of a negatively charged one, $\Xi^-$, was predicted by the Gell-Mann--Nishijima strangeness theory\tcite{Gell-Mann:1956iqa,Nishijima:1955gxk}.
The $\Xi^-$ baryon was discovered in the context of cosmic-ray experiments in 1952\tcite{Armenteros:1953wpa,Butler:1957nds,Armenteros:1987nt}. The $\Xi^0$ hyperon was observed for the first time at the Lawrence Berkeley Laboratory in 1959\tcite{Alvarez:1959zz}, and then as a daughter product for $\Omega^-$ baryon decays at the Brookhaven National Laboratory in 1964\tcite{Barnes:1964pd}.
Both the $\Xi^0$ and the $\Xi^+$ hyperons are part of the baryon \emph{octet}\tcite{Gell-Mann:1961omu,Neeman:1961jhl}.

By studying distributions differential in rapidity, azimuthal angle and/or transverse momenta, we will provide arguments supporting the statement that the $\Xi^-/\bar\Xi^+$ collinear FFs act as stabilizers of the high-energy series. The found stabilization effects are milder than the ones generated by FFs depicting heavy-flavored hadron productions, but enough relevant to allow us for a study of our distributions at the natural energy scales indicated by kinematics.
We will calculate these observables within the full $\NLL$ accuracy by considering two different representations for resummed cross sections.
One of these representations contains terms which genuinely go beyond the NLL level. 
Thus, we will provide the first and pioneer systematic analysis on assessing effects coming from two distinct higher-order cross-section representations and from the inclusion of next-to-NLL contributions in a hadron-plus-jet hadroproduction process.

The structure of this article is the following. In Section\tref{sec:theory} we introduce the hybrid high-energy/collinear factorization and the observables of interest (Section\tref{ssec:kinematics}). In Section\tref{sec:results}, we present our phenomenological analysis, after giving highlights on the stabilization mechanism connected to fragmentation (Section\tref{ssec:stability}). Finally (Section~\ref{sec:conclusions}), we come out with conclusions and outlook.

%-----------------------------------------
\section{Hybrid factorization at work}
\label{sec:theory}
%-----------------------------------------

The reaction matter of our analysis is (see Fig.\tref{fig:process})
\begin{equation}
\label{process}
    {\rm p}(P_a) + {\rm p}(P_b) \, \rightarrow \, \Xi(\kappa_\Xi, \varphi_\Xi, y_\Xi) + {\cal X} + {\rm jet}(\kappa_J,\varphi_J, y_J) \; ,
\end{equation}
where ${\rm p}(P_{a,b})$ is a parent proton with momentum $P_{a,b}$, $\Xi(\kappa_\Xi, y_\Xi)$ is a $\Xi^-$ baryon, or its antiparticle $\bar\Xi^+$, detected with momentum $\kappa_\Xi$, azimuthal angle $\varphi_\Xi$ and rapidity $y_\Xi$, the jet is emitted with momentum $\kappa_J$, azimuthal angle $\varphi_J$ and rapidity $y_J$, and ${\cal X}$ denotes all the undetected products. The large transverse momenta, $|\vec \kappa_{\Xi,J}|$, and the high rapidity separation, $\DY \equiv y_\Xi - y_J$, allows us to access diffractive semi-hard configurations in the final state.

The momenta of the two parent protons form a Sudakov-vector basis satisfying $P_a^2= P_b^2=0$ and $(P_a\cdot P_b) = s/2$, so that the outgoing-object momenta can be decomposed as
\begin{equation}\label{sudakov}
\kappa_{\Xi,J} = x_{\Xi,J} P_{a,b} + \frac{\vec \kappa_{\Xi,J}^{\,2}}{x_{\Xi,J} s}P_{b,a} + \kappa_{\Xi,J\perp} \ , \quad
\kappa_{\Xi,J\perp}^2=-\vec \kappa_{\Xi,J}^{\,2}\;.
\end{equation}
Here, the longitudinal momentum fractions of our final-state particles, $x_{\Xi,J}$, are connected to the corresponding rapidities by the relations
$y_{\Xi,J}=\pm\frac{1}{2}\ln\frac{x_{\Xi,J}^2 s}
{\vec \kappa_{\Xi,J}^2}$ and $\drv y_{\Xi,J} = \pm \frac{\drv x_{\Xi,J}}{x_{\Xi,J}}$. We have 
\begin{equation}
\label{DeltaY}
 \DY = y_\Xi - y_J = \ln \left( \frac{x_\Xi x_J}{|\vec \kappa_\Xi||\vec \kappa_J|} s \right) \;.
\end{equation}

%-----------------------------------------
\subsection{Differential cross section at NLL}
\label{ssec:cross_section}
%-----------------------------------------

In a pure QCD collinear-factorization approach at LO, the differential cross section for our reaction reads as a one-dimensional convolution between on-shell parton hard factor, the parent-proton PDFs, and baryon FFs
\begin{equation}
\label{sigma_collinear}
\begin{split}
\frac{\drv\sigma^{\rm LO}_{\rm [coll.]}}{\drv x_\Xi\drv x_J\drv ^2\vec \kappa_\Xi\drv ^2\vec \kappa_J}
= \hspace{-0.25cm} \sum_{u,v=q,{\bar q},g}\int_0^1 \hspace{-0.20cm} \drv x_a \int_0^1 \hspace{-0.20cm} \drv x_b\ f_u\left(x_a\right) f_v\left(x_b\right)%\\
%\times
\int_{x_\Xi}^1 \hspace{-0.15cm} \frac{\drv \zeta}{\zeta}D^{\Xi}_{u}\left(\frac{x_\Xi}{\zeta}\right) 
\frac{\drv {\hat\sigma}_{u,v}\left(\hat s\right)}
{\drv x_\Xi\drv x_J\drv ^2\vec \kappa_\Xi\drv ^2\vec \kappa_J}\;.
\end{split}
\end{equation}
The $u,v$ indices stand for the parton species (quarks $q = u, d, s, c, b$; antiquarks $\bar q = \bar u, \bar d, \bar s, \bar c, \bar b$; gluon $g$), $f_{u,v}\left(x_{a,b}, \mu_F \right)$ are the proton PDFs and $D^{\Xi}_{u}\left(x_\Xi/\zeta, \mu_F \right)$ denote the $\Xi$-particle FFs; $x_{a,b}$ are the longitudinal fractions of the partons entering the hard subprocess, while $\zeta$ represents the longitudinal fraction of the outgoing parton fragmenting into $\Xi$. Then, $\drv\hat\sigma_{u,v}\left(\hat s \right)$ is the partonic hard factor, where $\hat s \equiv x_a x_b s$ is the squared center-of-mass energy of the partonic collision.
For the sake of simplicity, the explicit dependence factorization scale, $\mu_F$, has been everywhere dropped.

At variance with collinear factorization, the expression for the resummed cross section in our hybrid formalism is written in terms of a high-energy factorization, genuinely encoded in the BFKL formalism, between the Green's function and two forward-production impact factors. Collinear PDFs and FFs are then embodied via a one-dimensional convolution in the latters.
It is convenient to rewrite the differential cross section as a Fourier sum of azimuthal-angle coefficients
\begin{equation}
 \label{dsigma_Fourier}
 \frac{\drv \sigma}{\drv y_\Xi \drv y_J \drv \vec p_\Xi \drv \vec p_J \drv \varphi_\Xi \drv \varphi_J} =
 \frac{1}{(2\pi)^2} \left[{\cal C}_0 + 2 \sum_{m=1}^\infty \cos (m \varphi)\,
 {\cal C}_m \right]\, ,
\end{equation}
where $\varphi = \varphi_\Xi - \varphi_J - \pi$ contains the difference between final-state particles' azimuthal angles.

The first building block of the resummed cross section is the NLL Greens' function

\begin{equation}
\label{G_BFKL_NLL}
 {\cal G}_{\rm NLL}(\DY,m,\nu,\mu_R) = e^{{\DY} \bar \alpha_s(\mu_R) \,
 \chi^{\rm NLO}(m,\nu)} \; ,
\end{equation}
with $\bar \alpha_s(\mu_R) \equiv \alpha_s(\mu_R) N_c/\pi$, $N_c$ the number of colors, and $\beta_0 = 11N_c/3 - 2 n_f/3$ the first coefficient of the QCD $\beta$-function.
The BFKL kernel entering the exponent of Eq.~\eqref{G_BFKL_NLL} contains the NLL resummation of energy logarithms
\begin{eqnarray}
 \label{chi}
 \chi^{\rm NLO}(m,\nu) = \chi(m,\nu) + \bar\alpha_s \tilde{\chi}(m,\nu) \;,
\end{eqnarray}
where $\chi(m,\nu)$ stand for the eigenvalues of the kernel at LO
\begin{eqnarray}
 \label{kernel_LO}
 \chi\left(m,\nu\right) = -2\gamma_{\rm E} - 2 \, {\rm Re} \left\{ \psi\left(\frac{m+1}{2} + i \nu \right) \right\} \, ,
\end{eqnarray}
where $\gamma_{\rm E}$ is the Euler-Mascheroni constant and $\psi(z) \equiv \Gamma^\prime
(z)/\Gamma(z)$ the logarithmic derivative of the Gamma function. The $\tilde{\chi}(m,\nu)$ function in Eq.\eref{chi} is the NLO BFKL kernel correction
\begin{equation}
\label{chi_NLO}
\tilde{\chi} \left(m,\nu\right) = \bar\chi(m,\nu)+\frac{\beta_0}{8 N_c}\chi(m,\nu)
%\\ \nonumber 
%\times %\,
\left\{ -\chi(m,\nu) + \frac{10}{3} + 4\ln\frac{\mu_R}{\tilde{\mu}} \right\} \;,
\end{equation}
with the characteristic $\bar\chi(m,\nu)$ function calculated in Refs.~\cite{Kotikov:2000pm,Kotikov:2002ab}. Then, $\tilde{\mu} = \sqrt{M_{\Xi \perp} |\vec \kappa_J|}$, with $M_{\Xi \perp} = \sqrt{M_\Xi^2 + |\vec \kappa_\Xi|^2}$ the transverse mass of the $\Xi$ baryon and $M_\Xi = 1.32171$~GeV its mass.

The second building block is the forward-hadron NLO impact factor, calculated in the Mellin space by the projection onto the LO BFKL eigenfunctions. We rely on the calculation done in Ref.\tcite{Ivanov:2012iv}, which is suited for light-flavored bound states as well as heavy-flavored ones detected at large transverse momenta.
Its expression reads
\begin{equation}
\label{HIF}
\Phi_\Xi^{\rm NLO}(m,\nu,|\vec \kappa|,x) =
\Phi_\Xi(\nu,|\vec \kappa|,x) +
\alpha_s(\mu_R) \, \hat \Phi_\Xi(m,\nu,|\vec \kappa|,x) \;,
\end{equation}
where the LO part is given by
\begin{equation}
\label{LOHIF}
\hspace{-0.30cm}
\Phi_\Xi(\nu,|\vec \kappa|,x) 
= 2 \sqrt{\frac{C_F}{C_A}}
|\vec \kappa|^{2i\nu-1}
\!\!\int_{x}^1\frac{\drv \zeta}{\zeta}
\left( \frac{\zeta}{x} \right)
^{2 i\nu-1} 
%\\ \nonumber
%\times 
 \left[\frac{C_A}{C_F}f_g(\zeta)D_g^\Xi\left(\frac{x}{\zeta}\right)
 +\sum_{u=q,\bar q}f_u(\zeta)D_u^\Xi\left(\frac{x}{\zeta}\right)\right] \;,
\end{equation}
while the NLO correction, $\hat \Phi_\Xi(m,\nu,|\vec \kappa|,x)$, can be found in Eqs.~(4.58) to~(4.65) of Ref.\tcite{Ivanov:2012iv} (see also Eqs.~(77) to~(84) of its open-access arXiv version).

The last ingredient is the forward-jet NLO impact factor in the Mellin representation
\begin{equation}
\label{JIF}
\Phi_J^{\rm NLO}(m,\nu,|\vec \kappa|,x) =
\Phi_J(\nu,|\vec \kappa|,x) +
\alpha_s(\mu_R) \, \hat \Phi_J(m,\nu,|\vec \kappa|,x) \;,
\end{equation}
with
\begin{equation}
 \label{LOJIF}
 \Phi_J(\nu,|\vec \kappa|,x) =  2 \sqrt{\frac{C_F}{C_A}}
 |\vec \kappa|^{2i\nu-1}\left[\frac{C_A}{C_F}f_g(x)
 +\sum_{v=q,\bar q}f_v(x)\right] \;
\end{equation}
the LO contribution.
The expression for the NLO correction depends on the jet algorithm. We employ the formula obtained by combining Eq.~(36) of Ref.~\cite{Caporale:2012ih} with Eqs.~(4.19) and~(4.20) of Ref.~\cite{Colferai:2015zfa} (Eqs.~(53) and~(54) of its open-access arXiv version).
It relies on computations performed in Refs.\tcite{Ivanov:2012iv,Ivanov:2012ms}, suited to numerical analyses, where a jet algorithm calculated in the ``small-cone'' approximation (SCA)~\cite{Furman:1981kf,Aversa:1988vb} is adopted in the cone-type case (for further details, see Ref.~\cite{Colferai:2015zfa}). Following the choice done in recent CMS experimental studies on Mueller--Navelet jets, we fix the jet-cone radius to $R_J = 0.5$\tcite{Khachatryan:2016udy}.

Combining all the ingredients, we come out with a consistent definition of NLL-resummed azimuthal coefficients, valid in the $\MSb$ renormalization scheme\tcite{PhysRevD.18.3998}. We write (for technical details, see Ref.~\cite{Caporale:2012ih})
\begin{eqnarray}
\label{Cm_NLLp_MSb}%\nonumber
 \CmNLLp \!\! &=& \!\! %\frac{x_1 x_2}{|\vec \kappa_\Xi| |\vec \kappa_J|} 
 \frac{e^{\DY}}{s} 
 \int_{-\infty}^{+\infty} \drv \nu \, 
 %e^{{\DY} \bar \alpha_s(\mu_R) \chi^{\rm NLO}(m,\nu)} 
 {\cal G}_{\rm NLL}(\DY,m,\nu,\mu_R) \,
 \alpha_s^2(\mu_R) %\,
% \end{equation}
%\[
 \\ \nonumber
% &&
 \!\! &\times& \!\! \biggl\{\Phi_\Xi^{\rm NLO}(m,\nu,|\vec \kappa_\Xi|, x_\Xi)[\Phi_J^{\rm NLO}(m,\nu,|\vec \kappa_J|,x_J)]^*
 \\ \nonumber
 \!\! &+& \!\!
 \left.
 \alpha_s^2(\mu_R) \DY \frac{\beta_0}{4 \pi}
 \left[\frac{i}{2} \, \frac{\drv}{\drv \nu} \ln\frac{\Phi_\Xi}{\Phi_J^*} + \ln\left(|\vec \kappa_\Xi| |\vec \kappa_J|\right)\right]
 \right\}
 \;.
% \]
\end{eqnarray}
The $\NLLp$ label indicates that a full NLL resummation of energy logarithms is consistently performed within the NLO perturbative accuracy.
The `$+$' superscript reflects that, in our representation for azimuthal coefficients, terms beyond the NLL level are generated by the cross product of the NLO impact-factor corrections.
Another representation, valid within the NLL accuracy and labeled as $\NLL$ is the one obtained by discarding the next-to-NLL factor coming from the cross product. 
In our analysis we will consider both the $\NLLp$ and $\NLL$ representations.
We will show that, for the considered final-state kinematics, the effect of switching from one to the other produces no relevant effects.

A comprehensive high-energy versus DGLAP study would rely on comparing observables calculated via our hybrid factorization and pure fixed-order computations.
According to our knowledge, however, a numerical code to study NLO distributions for inclusive semi-hard hadron-plus-jet hadroproductions is not yet available.
Thus, to assess the weight of the high-energy resummation on top of DGLAP predictions, we will compare our BFKL-inspired results with corresponding ones calculated by a high-energy fixed-order treatment, originally developed in the context of light di-jet\tcite{Celiberto:2015yba,Celiberto:2015mpa} and hadron-jet\tcite{Celiberto:2020wpk} azimuthal correlations.
It consists in truncating the high-energy series up to the NLO accuracy. This permits us to mimic the high-energy signal coming from a pure NLO calculation.
Operationally, we cut the expansion of azimuthal coefficients in Eq.~(\ref{Cm_NLLp_MSb}) up to ${\cal O}(\alpha_s^3)$. Thus, we obtain an effective high-energy fixed-order ($\HENLOp$) expression which can be easily adopted in our phenomenological study.
%It encodes leading-power asymptotic signal present in a pure NLO DGLAP calculation, and does not contains those factors which are suppressed by inverse powers of the partonic center-of-mass energy.
The $\MSb$ expression of the azimuthal coefficients in the $\HENLOp$ limit reads
\begin{eqnarray}
\label{Cn_HENLOp_MSb}%\nonumber
 \CmHENLOp \!\! &=& \!\! %\frac{x_1 x_2}{|\vec \kappa_\Xi| |\vec \kappa_J|} 
 \frac{e^{\DY}}{s} 
 \int_{-\infty}^{+\infty} \drv \nu \, 
 %e^{{\DY} \bar \alpha_s(\mu_R) \chi^{\rm NLO}(m,\nu)}
 \alpha_s^2(\mu_R) \,
 \big\{ {\cal G}_{\rm NLL}^{(0)}(\DY,m,\nu,\mu_R)
% \end{equation}
%\[
 \\ \nonumber
% &&
 \!\! &+& \!\!
 \Phi_\Xi^{\rm NLO}(m,\nu,|\vec \kappa_\Xi|, x_\Xi)[\Phi_J^{\rm NLO}(m,\nu,|\vec \kappa_J|,x_J)]^* \big\} \;,
% \]
\end{eqnarray}
with
\begin{equation}
\label{G_BFKL_0}
 {\cal G}_{\rm NLL}^{(0)}(\DY,m,\nu,\mu_R) = \bar \alpha_s(\mu_R) \DY \chi(m,\nu)
\end{equation}
the first term of the expansion of the BFKL Green's function in $\alpha_s$.
Analogously to the NLL case, it is possible to obtain a $\HENLO$ expression by removing the next-to-NLL factor coming from the cross product of the NLO corrections of the two impact factors.

We will also compare or BFKL and high-energy DGLAP predictions with corresponding ones taken in the pure LL limit, given in the $\MSb$ scheme by
\begin{equation}
\label{Cm_LL_MSb}%\nonumber
 \CmLL = %\frac{x_1 x_2}{|\vec \kappa_\Xi| |\vec \kappa_J|} 
 \frac{e^{\DY}}{s} 
 \int_{-\infty}^{+\infty} \drv \nu \, 
 e^{{\cal G}_{\rm NLL}^{(0)}(\DY,m,\nu,\mu_R)} %\,
 \alpha_s^2(\mu_R) \, \Phi_\Xi(m,\nu,|\vec \kappa_\Xi|, x_\Xi)[\Phi_J(m,\nu,|\vec \kappa_J|,x_J)]^* \;,
\end{equation}
with $\Phi_{\Xi,J}(m,\nu,|\vec \kappa_{\Xi,J}|, x_{\Xi,J})$ the LO $\Xi$-baryon and jet impact factors presented in Eqs.\eref{LOHIF} and\eref{LOJIF}, respectively.

Renormalization and factorization scales will be set to the \emph{natural} energies provided by the given final state. One has $\mu_R = \mu_F \equiv \mu_N$, with $\mu_N = M_{\Xi \perp} + |\vec \kappa_J|$ the \emph{natural} reference scale of the process. To assess the weight of higher-order corrections, $\mu_F$ and $\mu_R$ scales will be varied around $\mu_N$ by a factor controlled by the $C_\mu$ parameter (see Section\tref{sec:results}).

%-----------------------------------------
\subsection{Observables and kinematics}
\label{ssec:kinematics}
%-----------------------------------------

The first observable considered in our study is the \emph{rapidity distribution}, namely the cross section differential in the rapidity interval, $\DY$. Its expression can be got by integrating the ${\cal C}_0$ azimuthal coefficient over transverse momenta and rapidities of the two outgoing objects, while $\DY$ is kept fixed. We have
\begin{equation}
 \label{DY_distribution}
 \frac{\drv \sigma}{\drv \DY} =
 \int_{\kappa_\Xi^{\rm min}}^{\kappa_\Xi^{\rm max}} \!\!\drv |\vec \kappa_\Xi|
 \int_{\kappa_J^{\rm min}}^{\kappa_J^{\rm max}} \!\!\drv |\vec \kappa_J|
% \int_{y_\Xi^{\rm min}}^{y_\Xi^{\rm max}} \drv y_\Xi
 \int_{\max(\DY + y_J^{\rm min}, \, y_\Xi^{\rm min})}^{\min(\DY + y_J^{\rm max}, \, y_\Xi^{\rm max})} \drv y_\Xi
% \int_{y_J^{\rm min}}^{y_J^{\rm max}} \drv y_J
 %\, \,
 %\delta (\DY - (y_\Xi - y_J))
 \, \,
 {\cal C}_0\left(|\vec \kappa_\Xi|, |\vec \kappa_J|, y_\Xi, y_J \right)
\Bigm \lvert_{y_J \;=\; y_\Xi - \DY}
 \; ,
\end{equation}
where a $\delta (\DY - (y_\Xi - y_J))$ delta has been used to remove the integration in $y_J$ and to set the extremes of integration in $y_\Xi$ accordingly.
The $\Xi$ hadron is reconstructed by the CMS barrel detector, thus having $|y_\Xi| < 2.4$.
As for its transverse-momentum window, we admit 10~GeV~$< |\vec \kappa_\Xi| <$~35~GeV.
The jet is always detected in its typical CMS ranges\tcite{Khachatryan:2016udy}, namely $|y_J| < 4.7$ and 35~GeV~$< |\vec \kappa_J| <$~60~GeV.
Employing disjoint windows for the transverse momenta of the two emitted objects helps to better disentangle pure high-energy imprints from the DGLAP background\tcite{Celiberto:2015yba,Celiberto:2015mpa,Celiberto:2020wpk}. It also quenches Sudakov logarithmic contaminations rising from almost back-to-back final states that would require the use of another appropriate resummation\tcite{Mueller:2012uf,Mueller:2013wwa,Marzani:2015oyb,Mueller:2015ael,Xiao:2018esv}.
Furthermore, it suppresses possible instabilities rising in next-to-leading calculations\tcite{Andersen:2001kta,Fontannaz:2001nq} as well as NLL violations of the energy-momentum conservation\tcite{Ducloue:2014koa}.

The second observable matter of our interest is the \emph{azimuthal distribution}, namely the normalized cross section differential both in $\DY$ and in the azimuthal-angle distance, $\varphi$
\begin{eqnarray}
 \label{phi_distribution}
 \frac{1}{\sigma} \frac{\drv \sigma}{\drv \DY \drv \varphi} =
 \frac{1}{\pi} \left[\frac{1}{2} + \sum_{m=1}^\infty \langle \cos(m \varphi) \rangle \cos (m \varphi) \right] \; ,
\end{eqnarray}
where the mean values of azimuthal-angle cosines can be calculated as ratios of azimuthal coefficients, $\langle \cos(m \varphi) \rangle \equiv C_m/C_0$. Here, $C_m$ stand for the integrated azimuthal coefficients, obtained by generalizing the phase-space integration in Eq.\eref{DY_distribution} the differential ${\cal C}_{m > 0}$ ones.
Originally proposed in the context of Mueller--Navelet jets\tcite{Marquet:2007xx,Vera:2007kn,Ducloue:2013hia}, the azimuthal distribution represents one of the most promising observables where to hunt for the high-energy QCD dynamics. Indeed, it embodies signals coming from all azimuthal modes, and not just from $C_0$ or from a single cosine $\langle \cos(m \varphi) \rangle$. Moreover, being differential in $\varphi$, it eases the comparison with data, since detectors hardly cover the whole $(2\pi)$ range.
The outcome of a quite recent investigation on Mueller--Navelet $\varphi$-distributions was the study of these distributions allows us $(i)$ to overcome the well-known problems rising in the description of light-flavored final states at natural energy scales and $(ii)$ to enhance the agreement with experimental data collected at 7~TeV~CMS\tcite{Khachatryan:2016udy}.
We will present predictions for $\Xi$-plus-jet azimuthal distributions in the same kinematic ranges proposed above, and for given values of $\DY$.

The third observable is the \emph{double differential transverse-momentum distribution}
\begin{equation}
\label{2pT_distribution}
 \frac{\drv \sigma}{\drv \DY \drv |\vec \kappa_\Xi| \drv |\vec \kappa_J|} =
% \int_{y_\Xi^{\rm min}}^{y_\Xi^{\rm max}} \drv y_\Xi
 \int_{\max(\DY + y_J^{\rm min}, \, y_\Xi^{\rm min})}^{\min(\DY + y_J^{\rm max}, \, y_\Xi^{\rm max})} \drv y_\Xi
% \int_{y_J^{\rm min}}^{y_J^{\rm max}} \drv y_J
 %\, \,
 %\delta (\DY - (y_\Xi - y_J))
 \, \,
 {\cal C}_0\left(|\vec \kappa_\Xi|, |\vec \kappa_J|, y_\Xi, y_J \right)
\Bigm \lvert_{y_J \;=\; y_\Xi - \DY}
 \; ,
\end{equation}
\emph{i.e.} the cross section differential in $\DY$ and in the observed-particle transverse momenta, which we allow to lie in the 10~GeV~$< |\vec \kappa_{\Xi,J}| <$~100~GeV range.
This distribution was recently proposed in the context of high-energy inclusive emissions of bottom-flavored hadrons plus light-flavored jets as a common basis to unveil the interplay among different kinds of resummation mechanisms.
Indeed, when the transverse momenta stay in wider windows, other regions contiguous to the  semi-hard one are accessed.
As an example, when the transverse momenta are high or their mutual separation is large, the size of DGLAP-type logarithms and of \emph{threshold} contaminations\tcite{Sterman:1986aj,Catani:1989ne,Catani:1996yz,Bonciani:2003nt,deFlorian:2005fzc,Ahrens:2009cxz,deFlorian:2012yg,Forte:2021wxe,Mukherjee:2006uu,Bolzoni:2006ky,Becher:2006nr,Becher:2007ty,Bonvini:2010tp,Ahmed:2014era,Banerjee:2018vvb,Duhr:2022cob,Shi:2021hwx,Wang:2022zdu} increases. This makes the description afforded by a pure high-energy approach inadequate.
Then, in the very-low transverse-momentum regime, enhanced $|\vec \kappa|$-logarithms entering the perturbative expansion are not caught neglected by BFKL. Furthermore, \emph{diffusion-pattern} effects~\cite{Bartels:1993du} (see also Refs.~\cite{Caporale:2013bva,Ross:2016zwl}) grow and grow up to prevent the convergence of the high-energy resummation.
The most powerful way to take into account those  logarithms relies in an all-order transverse-momentum (TM) resummation\tcite{Catani:2000vq,Bozzi:2005wk,Bozzi:2008bb,Catani:2010pd,Catani:2011kr,Catani:2013tia,Catani:2015vma,Duhr:2022yyp}.
TM-resummed distributions have been recently investigated for the hadroproduction of photon\tcite{Cieri:2015rqa,Alioli:2020qrd,Becher:2020ugp,Neumann:2021zkb}, Higgs\tcite{Ferrera:2016prr} and $W$-boson\tcite{Ju:2021lah} pairs, and for boson-plus-jet\tcite{Monni:2019yyr,Buonocore:2021akg} and $Z$-plus-photon\tcite{Wiesemann:2020gbm} final states.
TM-based third-order fiducial predictions for Drell--Yan and Higgs emissions were presented in Refs.\tcite{Ebert:2020dfc,Re:2021con,Chen:2022cgv,Neumann:2022lft} and\tcite{Bizon:2017rah,Billis:2021ecs,Re:2021con,Caola:2022ayt}, respectively.
Finally, when the transverse momenta of the two detected particles lead to almost back-to-back final-state configurations, the previously mentioned Sudakov-type logarithms emerges and they need to be resummed as well\tcite{Mueller:2012uf,Mueller:2013wwa,Marzani:2015oyb,Mueller:2015ael,Xiao:2018esv}.
We will present predictions for $\Xi$-plus-jet double differential $|\vec \kappa|$-distributions without pretending to catch all the dominant features underlying these observables by the hands of our high-energy/collinear setup, but rather to set the ground for futures analyses aimed at unraveling the interplay among all these resummations.

We complement our study on the transverse-momentum spectrum of $\Xi$-baryon emissions by investigating the behavior of the $\kappa_\Xi$-distribution
\begin{equation}
\label{pT_distribution}
%\hspace{-0.25cm}
 \frac{\drv \sigma}{\drv |\vec \kappa_\Xi|} =
 \int_{\kappa_J^{\rm min}}^{\kappa_J^{\rm max}} \!\! \drv |\vec \kappa_J|
% \int_{y_\Xi^{\rm min}}^{y_\Xi^{\rm max}} \drv y_\Xi
 \int_{\DY^{\rm min}}^{\DY^{\rm max}} \!\! \drv \DY
 \int_{\max(\DY + y_J^{\rm min}, \, y_\Xi^{\rm min})}^{\min(\DY + y_J^{\rm max}, \, y_\Xi^{\rm max})} \!\!\! \drv y_\Xi
% \int_{y_J^{\rm min}}^{y_J^{\rm max}} \drv y_J
 %\, \,
 %\delta (\DY - (y_\Xi - y_J))
 \, \,
 {\cal C}_0\left(|\vec \kappa_\Xi|, |\vec \kappa_J|, y_\Xi, y_J \right)
\Bigm \lvert_{y_J \;=\; y_\Xi - \DY}
 \; ,
\end{equation}
namely the cross section differential in the $\Xi$-particle transverse momentum, and integrated in $\DY$ windows and in the 30~GeV~$< |\vec \kappa_J| <$~120~GeV jet transverse-momentum range.
We will highlight how this observable permits to clearly discriminate between BFKL and high-energy fixed-order predictions in the large-$|\vec \kappa_\Xi|$ regime.

%-----------------------------------------
\section{Phenomenology}
\label{sec:results}
%-----------------------------------------

The numerical analysis presented in this Section has been made by making use the {\tt JETHAD} multi-modular interface\tcite{Celiberto:2020wpk,Celiberto:2022rfj}.
The sensitivity of our observables on renormalization- and factorization-scale variations has been assessed by letting $\mu_{R,F}$ stay around the \emph{natural} values given by kinematics, up to a factor ranging from 1/2 to two, according to the $C_\mu$ scale parameter.
Uncertainty bands entering plots embodies the overall effect of scale variations and multi-dimensional integration over the final-state phase space. The latter has been steadily kept below 1\% by {\tt JETHAD} integrators.
Collinear PDFs are described via the novel {\tt NNPDF4.0} NLO determination~\cite{NNPDF:2021uiq,NNPDF:2021njg} as provided by the {\tt LHAPDF} package~\cite{Buckley:2014ana}.
It was obtained from global fits via the \emph{replica} method, originally proposed in Ref.\tcite{Forte:2002fg} in the context of neural-network techniques.
Collinear FFs employed in our analysis for the $\Xi^-/\bar\Xi^+$ octet baryons have been recently determined via the NLO {\tt SHKS22} fit\tcite{Soleymaninia:2022qjf} on data for single inclusive
electron-positron annihilations through the {\tt MontBlanc} neural-network framework\tcite{Khalek:2021gxf,Khalek:2022vgy} developed by the {\tt MAP} Collaboration\tcite{Bacchetta:2022awv} (see Ref.\tcite{Soleymaninia:2022alt} for a similar study on unidentified charged light-hadron FFs).
$\Lambda_c$~baryons and $\Lambda$~hyperons are depicted by {\tt KKSS19}\tcite{Kniehl:2020szu} and {\tt AKK08}\tcite{Albino:2008fy} NLO~FFs, respectively.
A two-loop running-coupling setup with $\alpha_s\left(M_Z\right)=0.118$ and a dynamic flavor number, $n_f$, is adopted.
All computations are done in the $\MSb$ scheme.
The center-of-mass energy is set to $\sqrt{s} = 14$ TeV.

%-----------------------------------------
\subsection{Natural stability}
\label{ssec:stability}
%-----------------------------------------

\begin{figure*}[!t]
\centering

   \includegraphics[scale=0.53,clip]{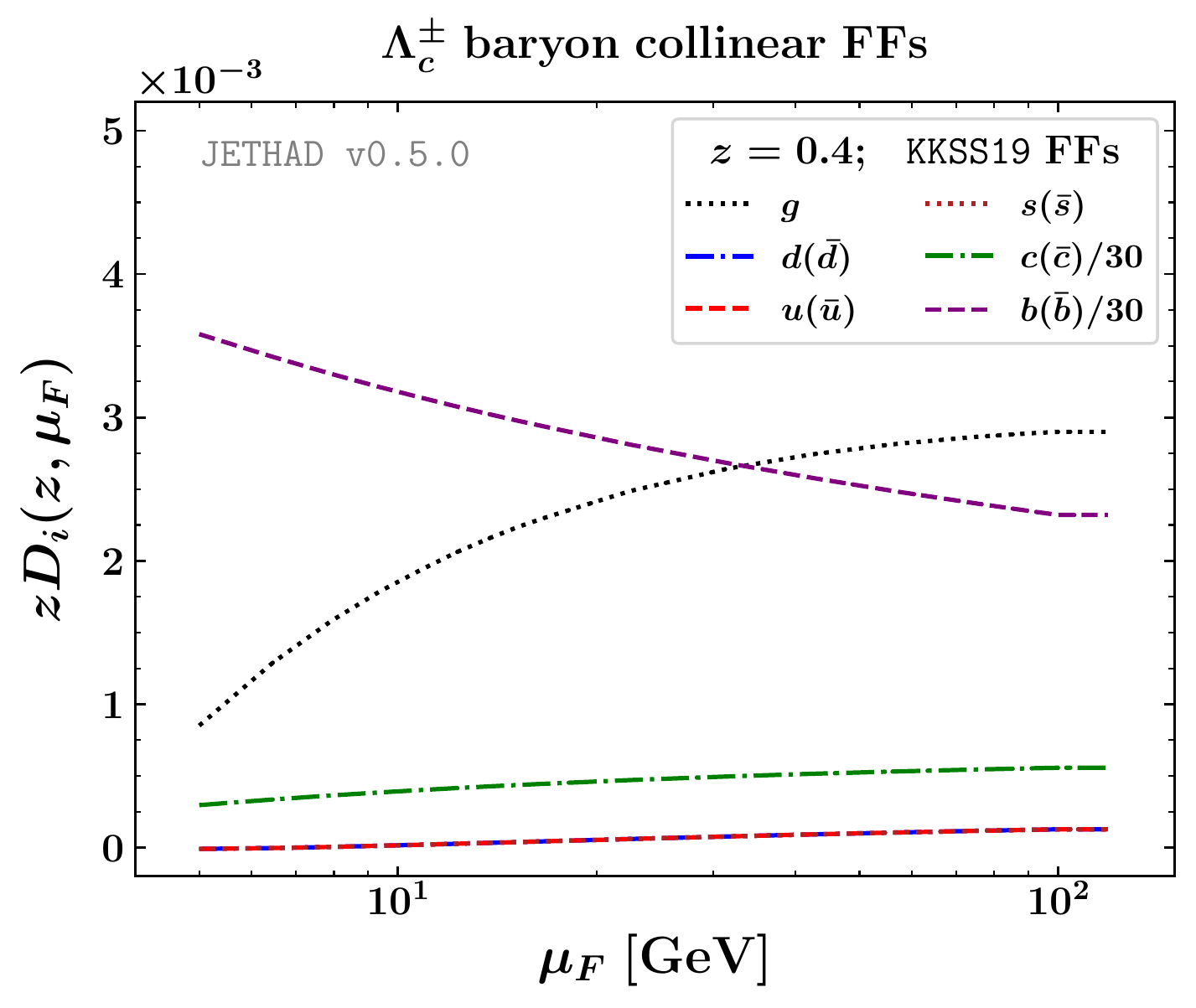}
%   \hspace{0.05cm}
   \includegraphics[scale=0.53,clip]{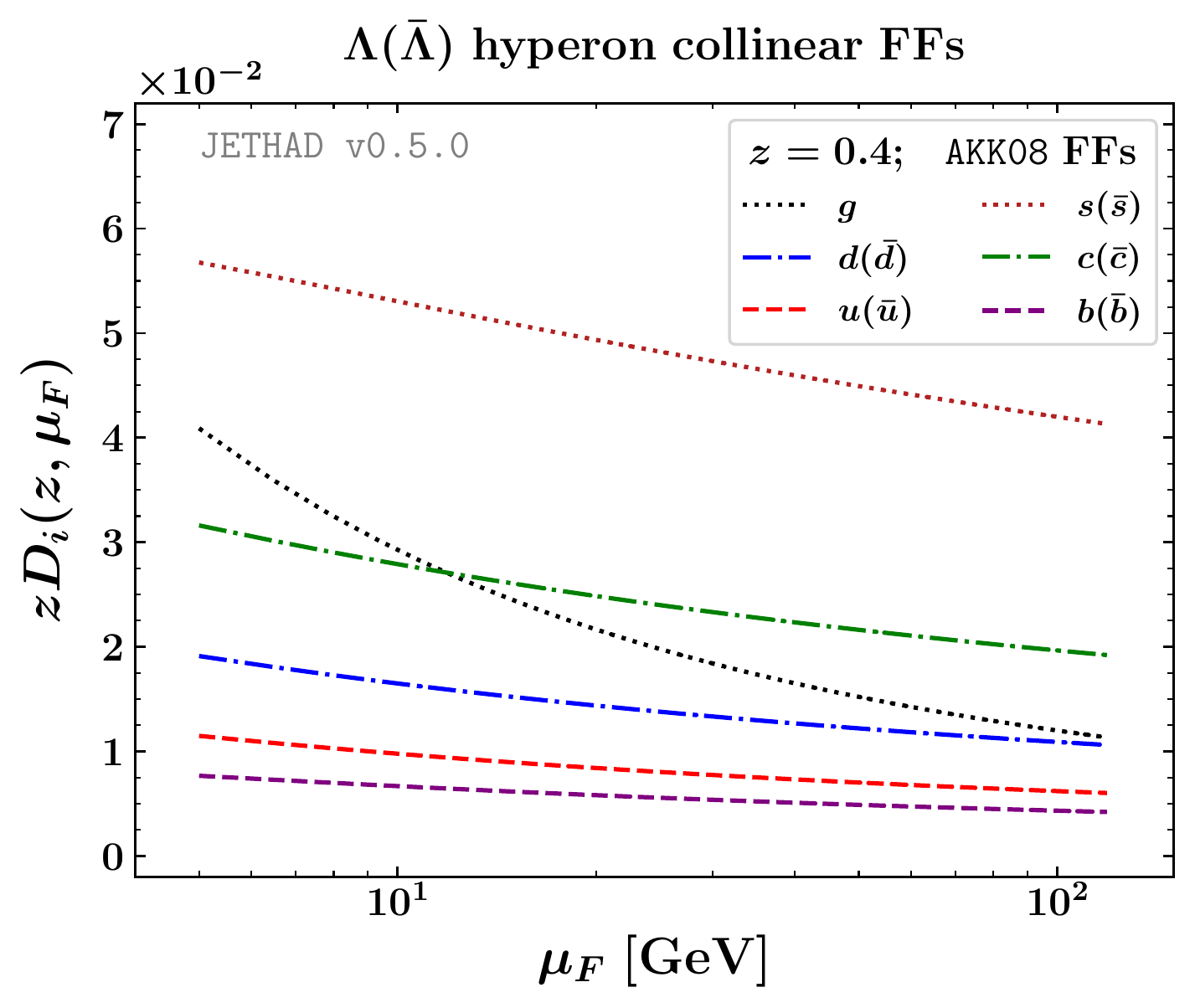}
%   \hspace{0.05cm}

   \includegraphics[scale=0.53,clip]{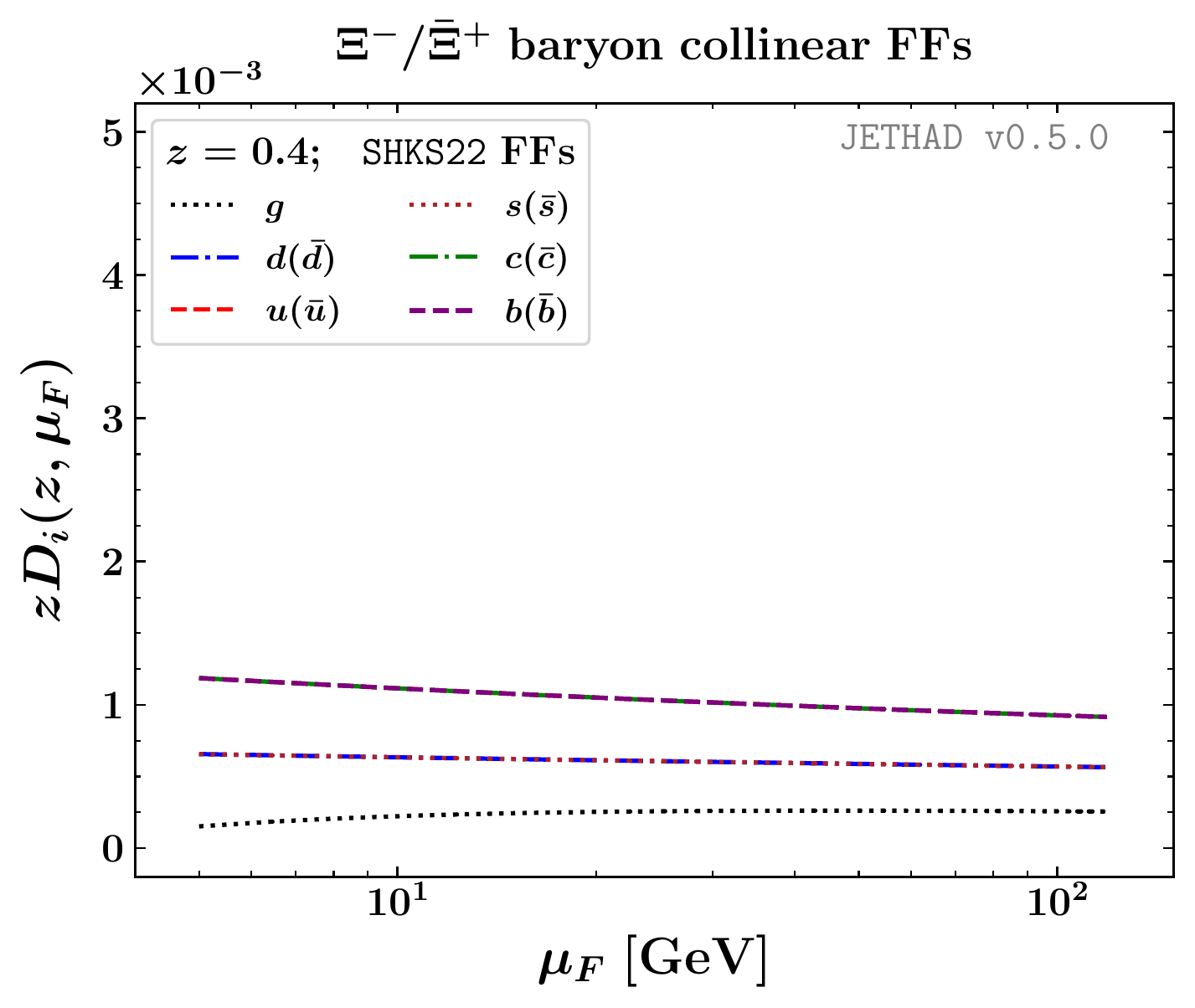}
%   \hspace{0.05cm}

\caption{Factorization-scale dependence of {\tt KKSS19}, {\tt AKK08}, and {\tt SHKS22} NLO FFs respectively depicting $\Lambda_c$-baryon, $\Lambda$-hyperon, and $\Xi$-baryon production, for $z = 0.4 \simeq \langle z \rangle$.}
\label{fig:FFs}
\end{figure*}

We present highlights on the stabilization mechanism emerging from the fragmentation mechanism depicting the production of bound states.
Details on the connection between the behavior of heavy-hadron VFNS FFs and the stability of high-energy resummed cross sections were recently discussed in Section~3.4 of Ref.\tcite{Celiberto:2021dzy} ($\Lambda_c$ baryons) and Appendix~A of Ref.\tcite{Celiberto:2021fdp} (noncharmed $B$ mesons and $\Lambda_b$ baryons).
In upper panels of Fig.\tref{fig:FFs} we consider the $\mu_F$-behavior of {\tt KKSS19} $\Lambda_c$ (left) and {\tt AKK08} $\Lambda$ (right) FF sets for a value of the hadron momentum fraction that roughly matches the average value of $z$ at which FFs are typically probed of our analysis, namely $z = 0.4 \simeq \langle z \rangle$. As expected, charm- and bottom-quark functions strongly prevail in $\Lambda_c$ production, while the strange-quark one prevails in $\Lambda$ fragmentation. 
We notice that the {\tt KKSS19} gluon function grows with $\mu_F$ up to reach a plateau. Conversely, the {\tt AKK08} gluon density falls off when $\mu_F$ increases. This dichotomy turns out to be relevant when FFs are diagonally convoluted with collinear PDFs in LO forward-hadron impact factors~(Eq.\eref{LOHIF}). In the kinematic sector of our interest, namely when in the $10^{-4} \lesssim x \lesssim 10^{-2}$, the gluon PDF heavily dominates over the quark channels, and the behavior of the gluon FF is enhanced.\footnote{As pointed out in Ref.\tcite{Celiberto:2021dzy}, this feature holds also at NLO, where $(qg)$ and $(gq)$ nondiagonal channels are opened (see Eq.~(4.58) of Ref.\tcite{Ivanov:2012iv}).} On one hand, larger $\mu_R$ values translate in a numerically smaller running coupling, both in the Green's function and in the impact factors. On the other hand, higher values of $\mu_F$ heighten the contribution of the gluon PDF. When the latter is convoluted with an increasing-with-$\mu_F$ gluon FF, such as the $\Lambda_c$ one, the two effects balance each other. This gives rise to the stability of $\Lambda_c$-distributions under scale variations. Conversely, the decreasing pattern of the $\Lambda$-hyperon gluon FF when $\mu_F$ increases prevents the two effects to offset each other. This hampers any possibility of reaching a stability in the description of $\Lambda$-sensitive high-energy cross sections.
In the lower panel of Fig.\tref{fig:FFs} we show the $\mu_F$-dependence of {\tt SHKS22} FFs depicting $\Xi$-baryon emissions at $z = 0.4$. We observe a smooth-behaved, nondecreasing-with-$\mu_F$ pattern of the gluon FF. It represents an intermediate situation between the $\Lambda_c$ and the $\Lambda$ case.
We will provide arguments supporting the statement that this peculiar behavior is responsible for a stabilization pattern of high-energy cross sections sensitive $\Xi$-baryon detections, weaker than what happens in the $\Lambda_c$ case, but still present.

%-----------------------------------------
\subsection{Rapidity distributions}
\label{ssec:C0}
%-----------------------------------------

\begin{figure*}[!t]
\centering

   \hspace{0.00cm}
   \includegraphics[scale=0.41,clip]{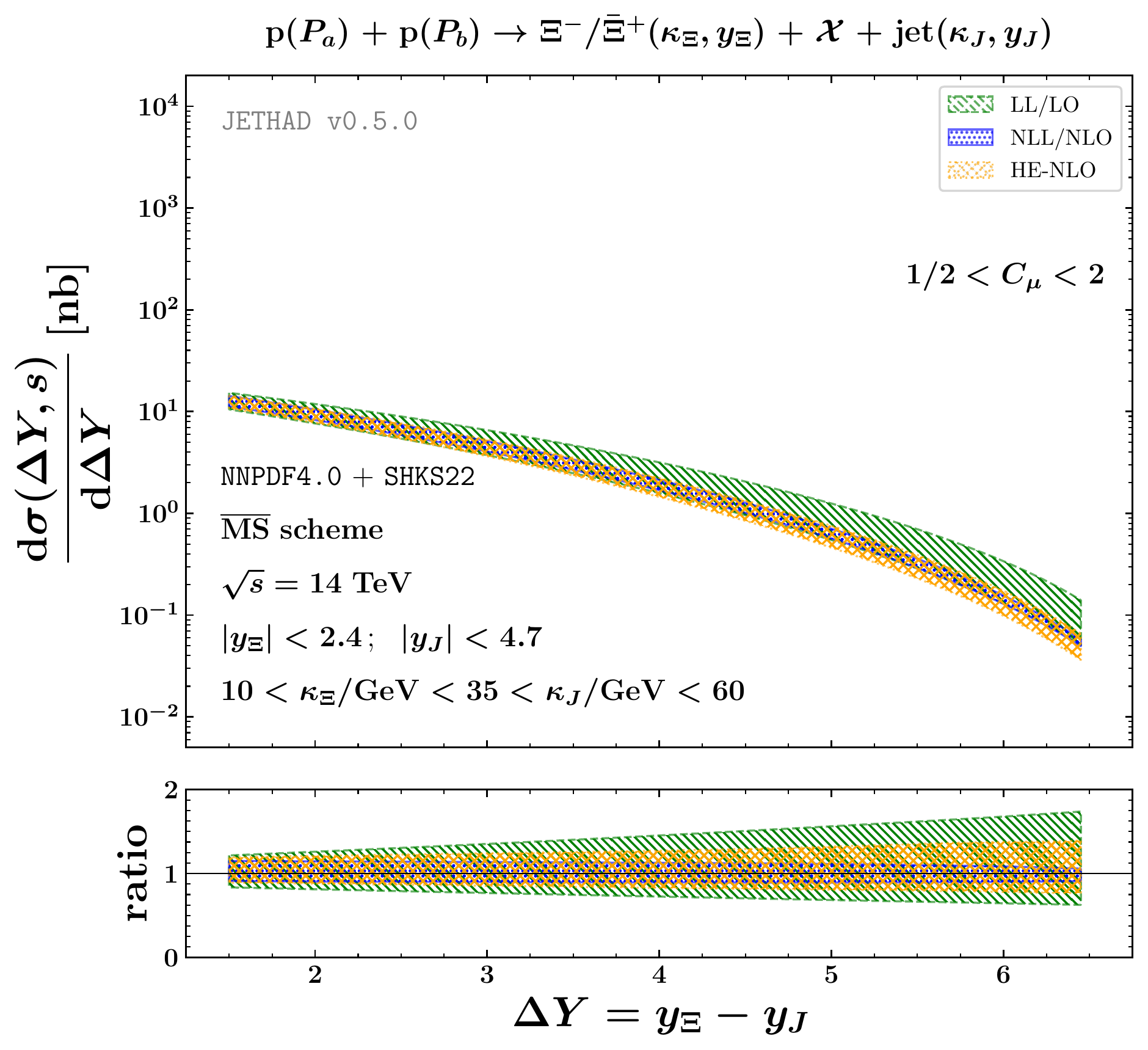}
   \hspace{-0.30cm}
   \includegraphics[scale=0.41,clip]{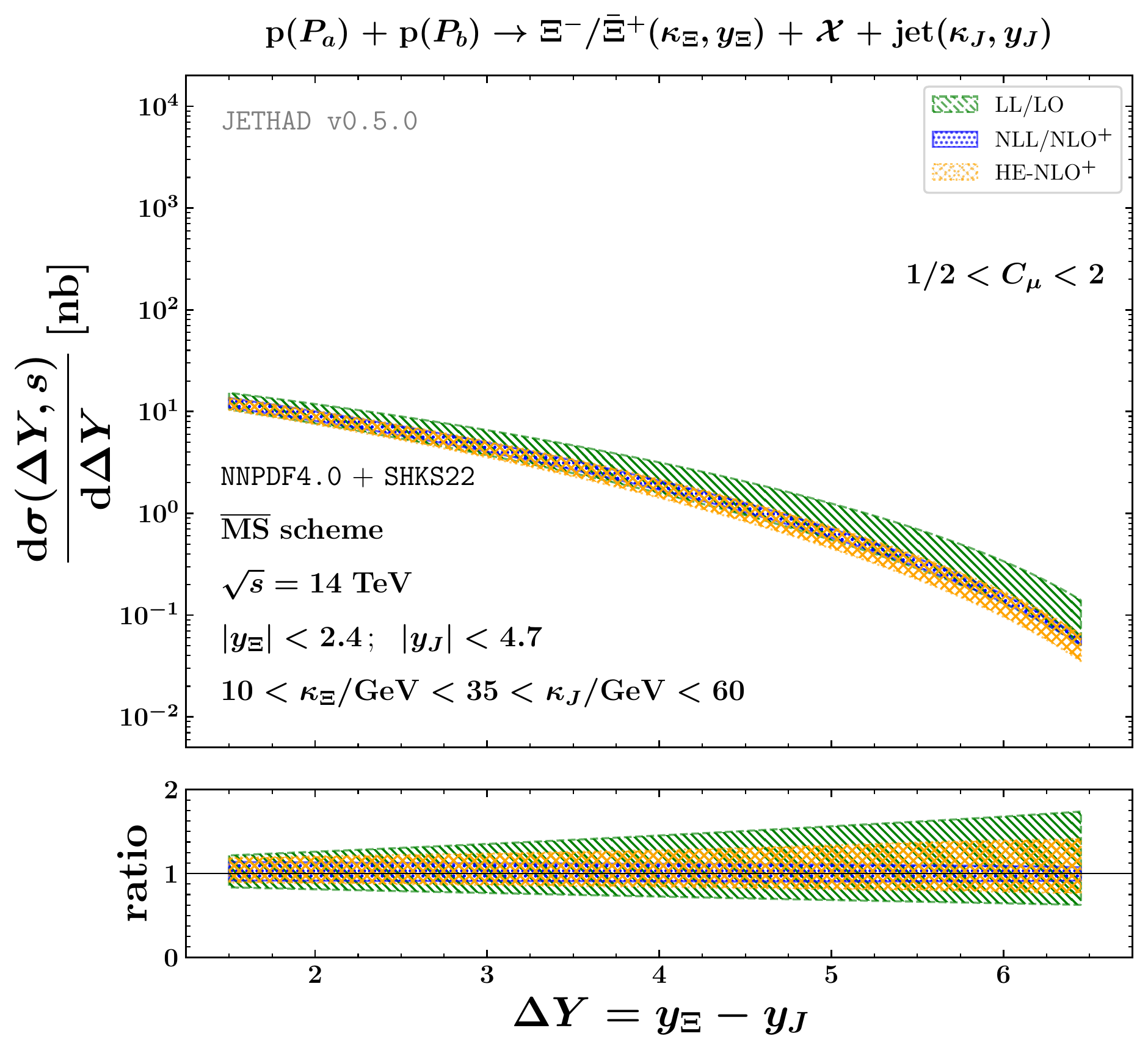}

   \includegraphics[scale=0.41,clip]{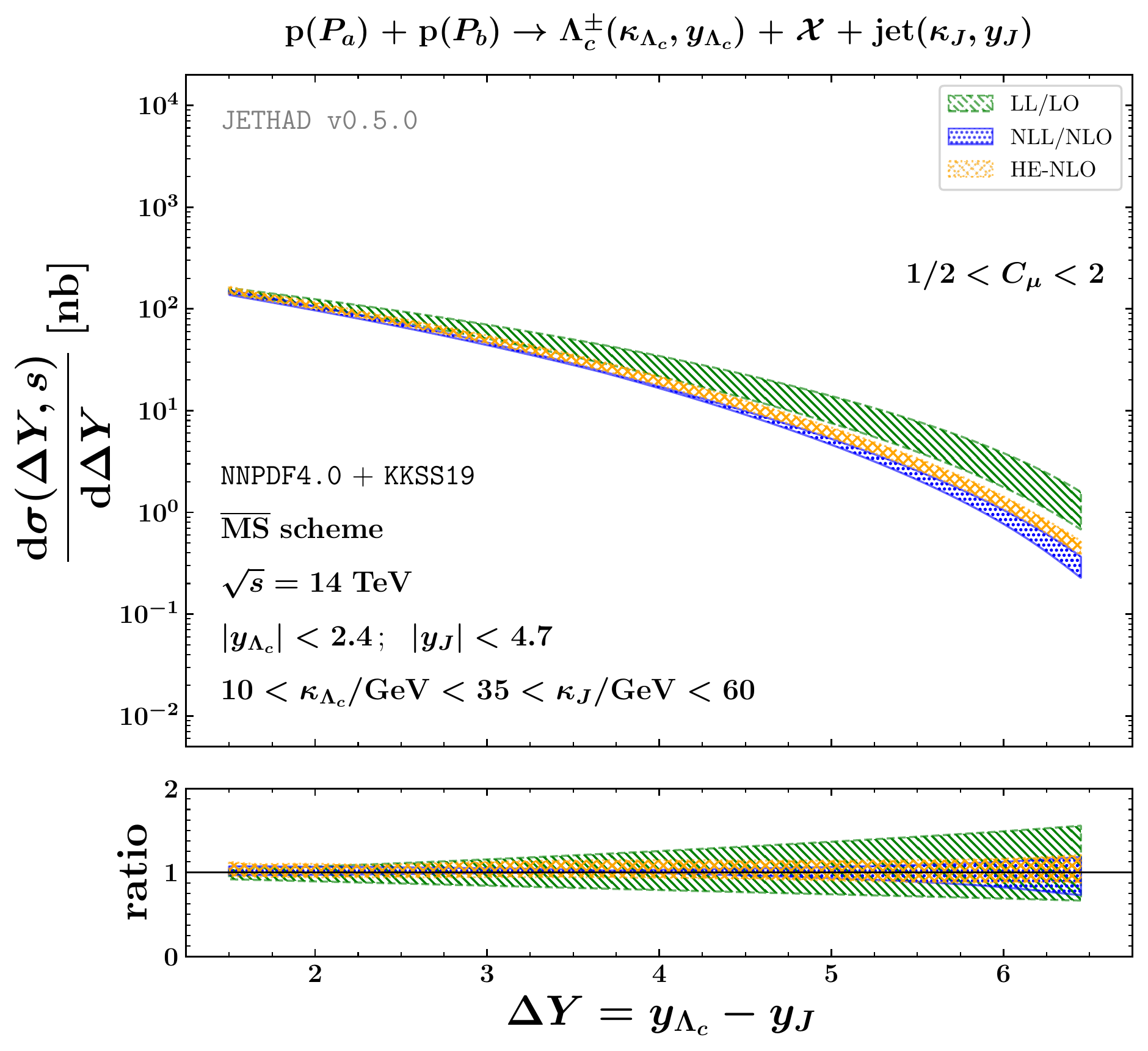}
%   \hspace{0.05cm}
   \includegraphics[scale=0.41,clip]{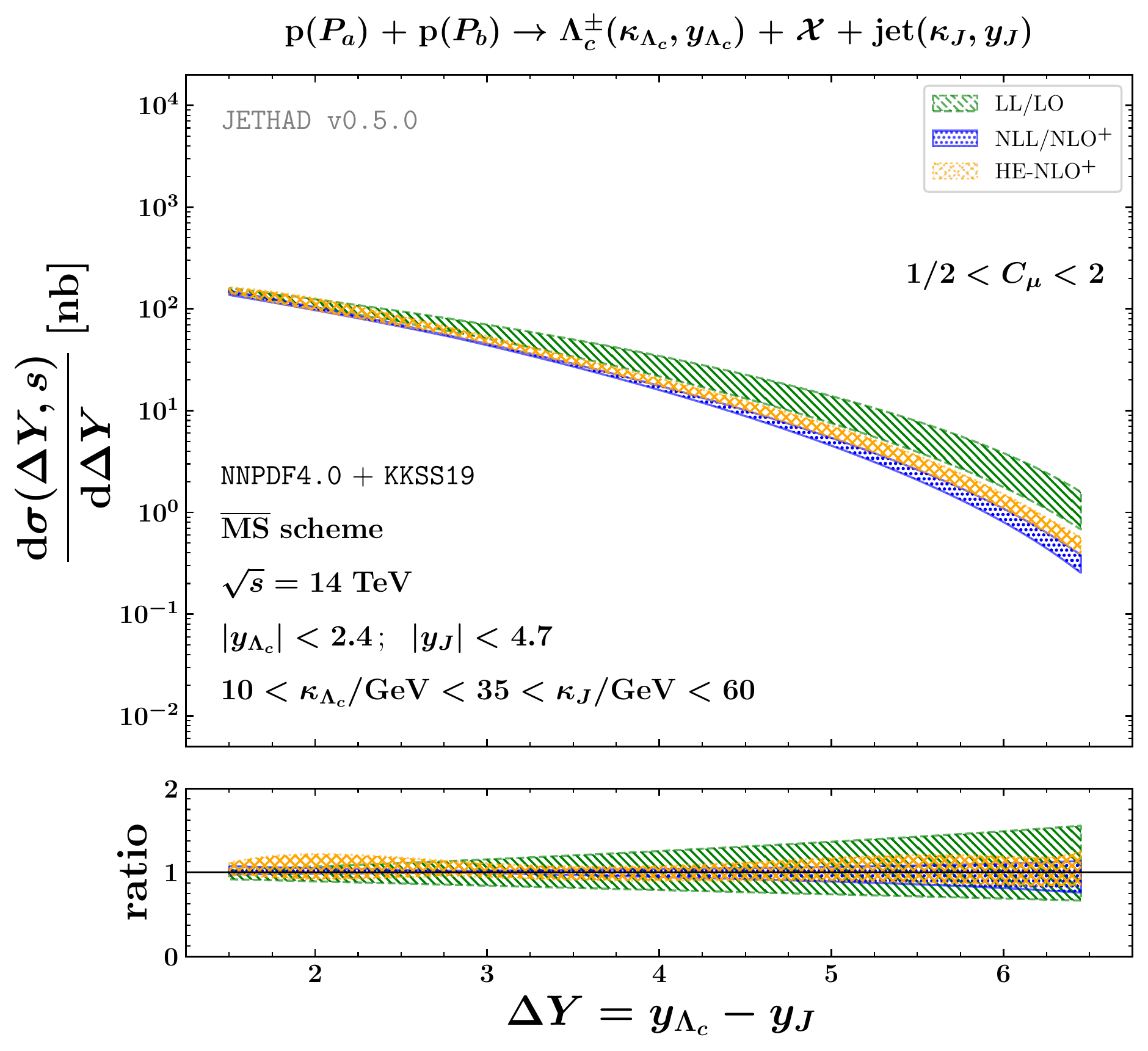}

   \includegraphics[scale=0.41,clip]{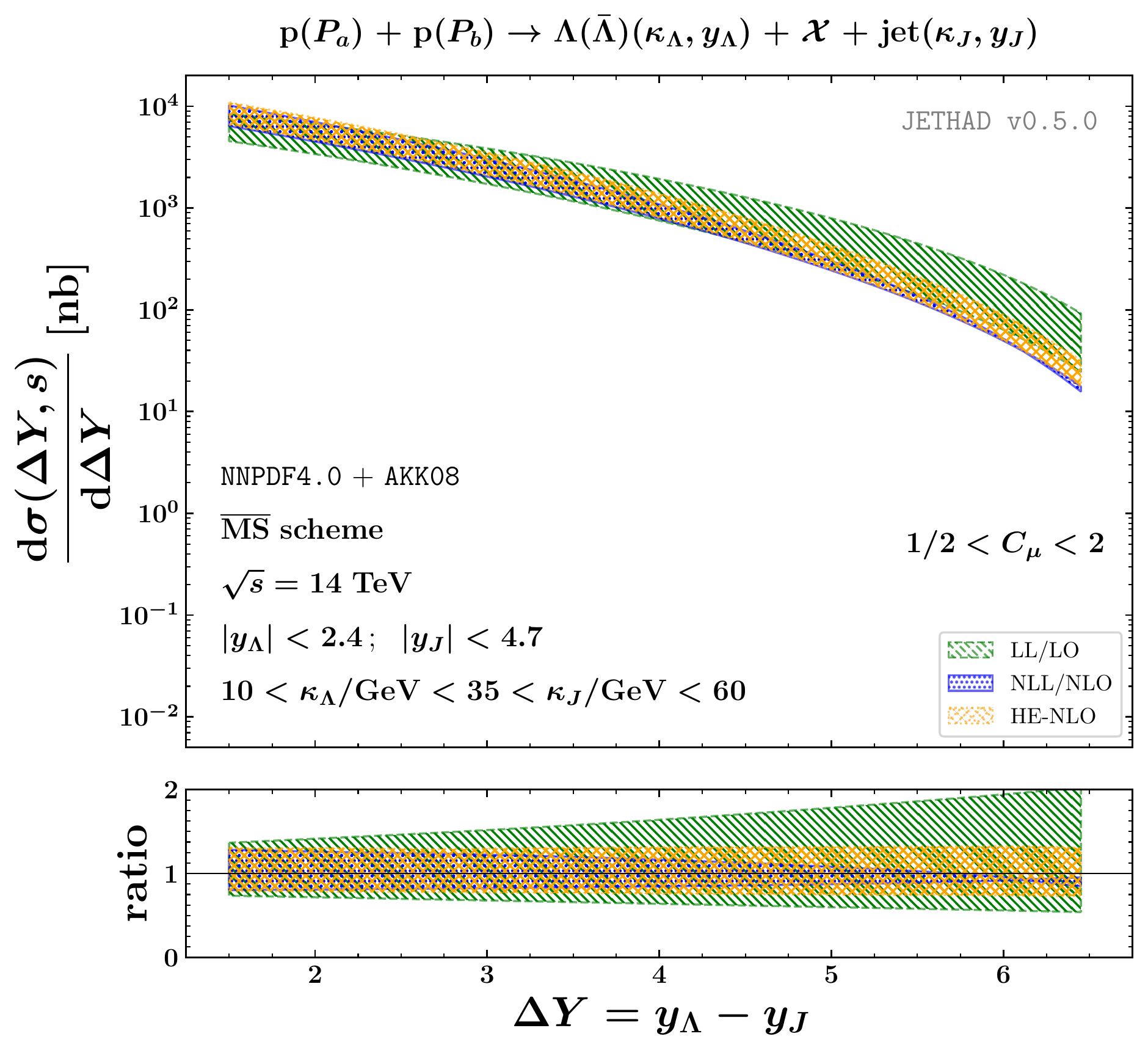}
%   \hspace{0.05cm}
   \includegraphics[scale=0.41,clip]{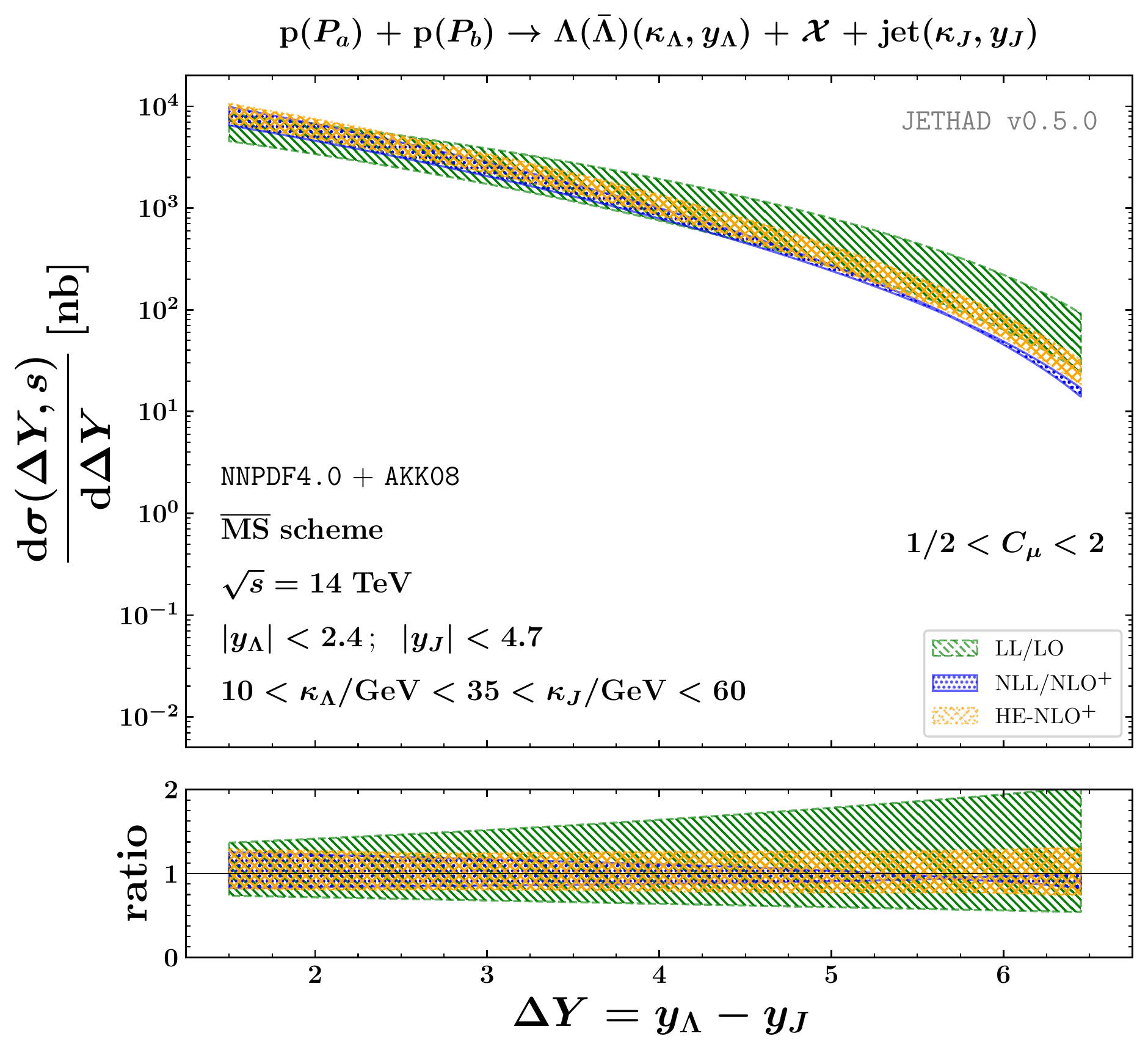}

\caption{$\DY$-dependence of the rapidity distribution within the $\NLL$ (left) and $\NLLp$ (right) accuracy, for $\Xi$-baryon (upper), $\Lambda_c$-baryon (central), and $\Lambda$-hyperon (lower) plus jet detections at $\sqrt{s} = 14$ TeV.
Shaded bands embody the combined effect of renormalization- and factorization-scale variation in the $1 < C_{\mu} < 2$ range and of phase-space numerical multidimensional integration.}
\label{fig:C0}
\end{figure*}

\begin{figure*}[!t]
\centering

   \hspace{0.00cm}
   \includegraphics[scale=0.41,clip]{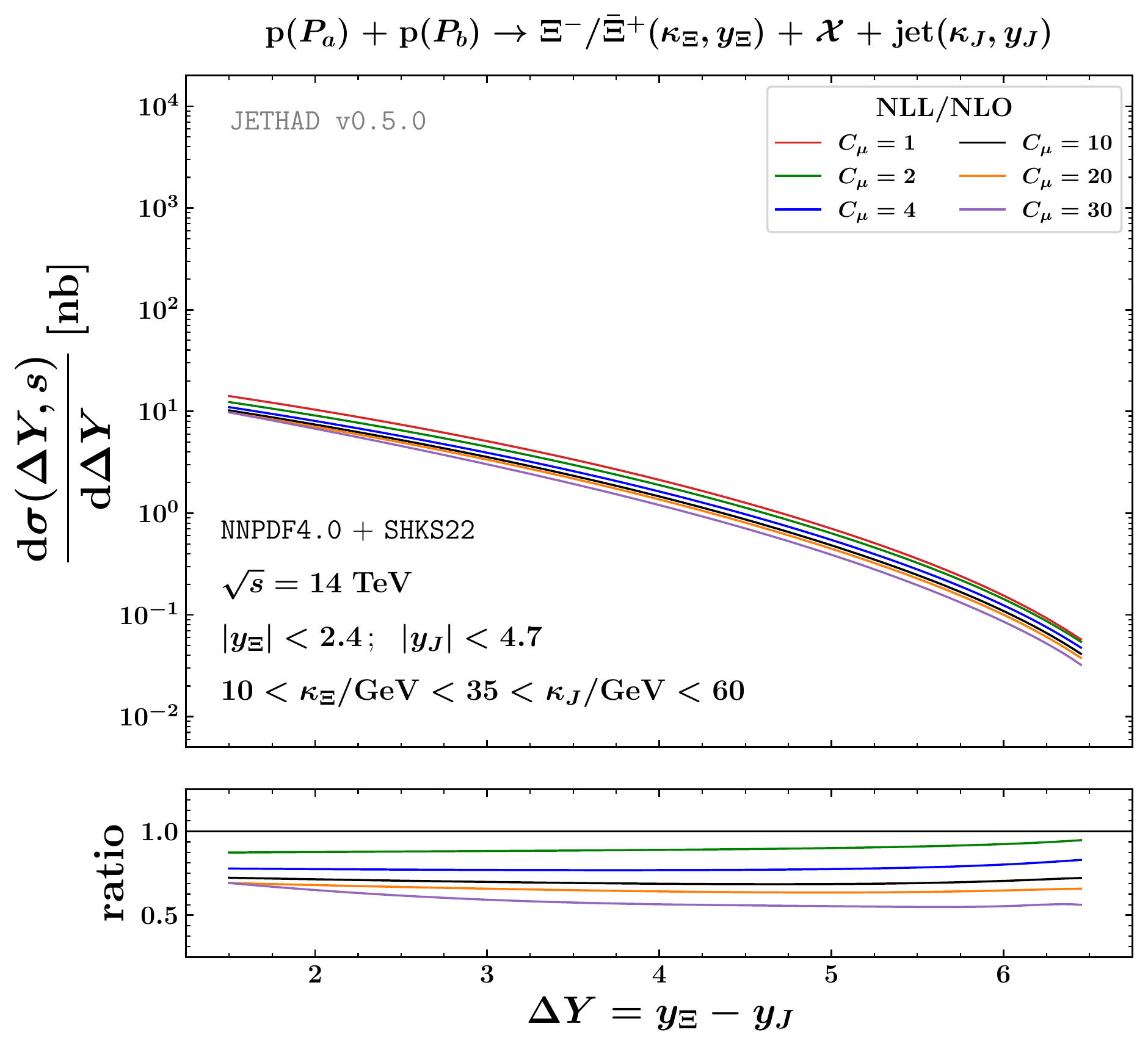}
   \hspace{-0.30cm}
   \includegraphics[scale=0.41,clip]{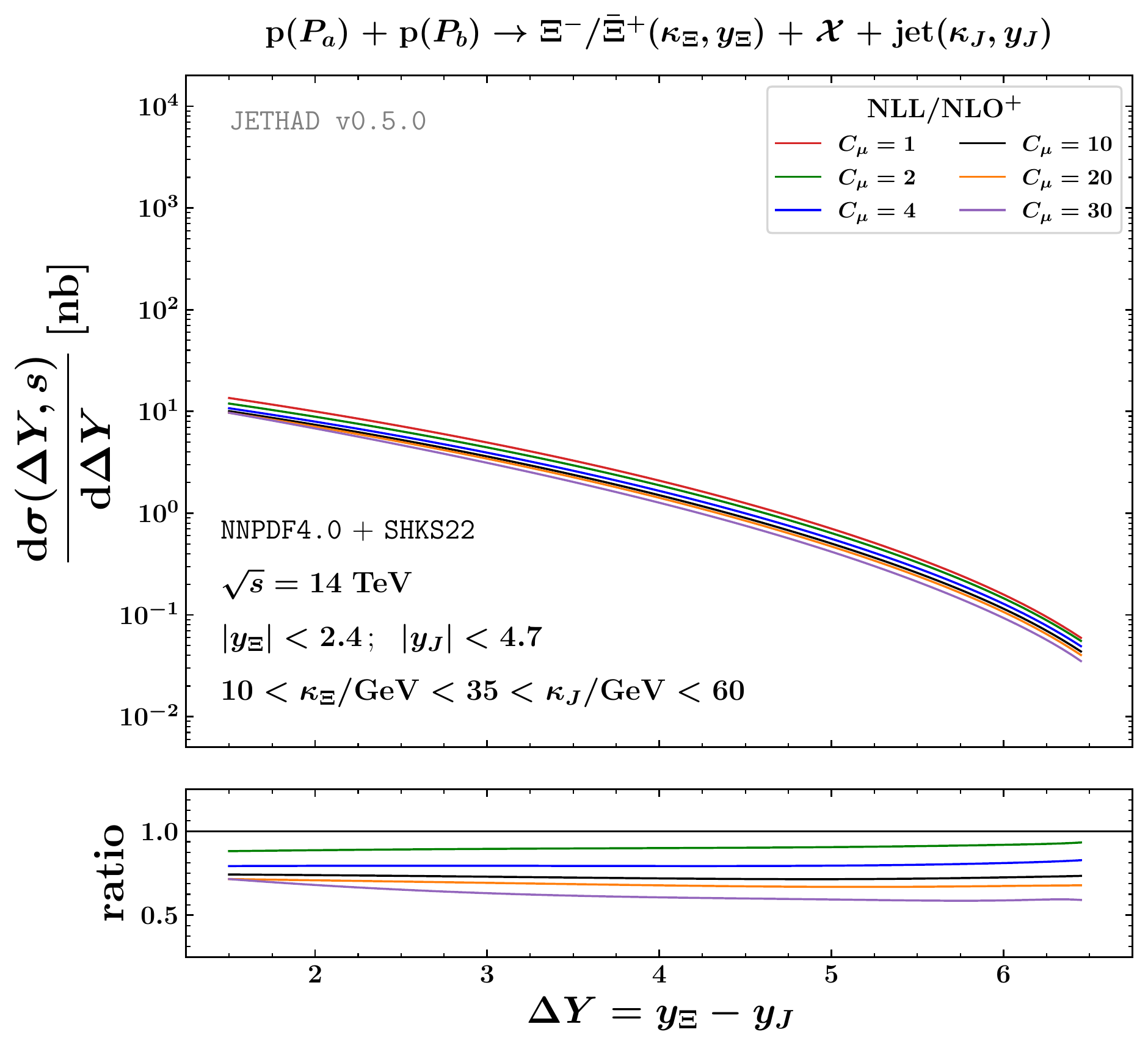}

   \includegraphics[scale=0.41,clip]{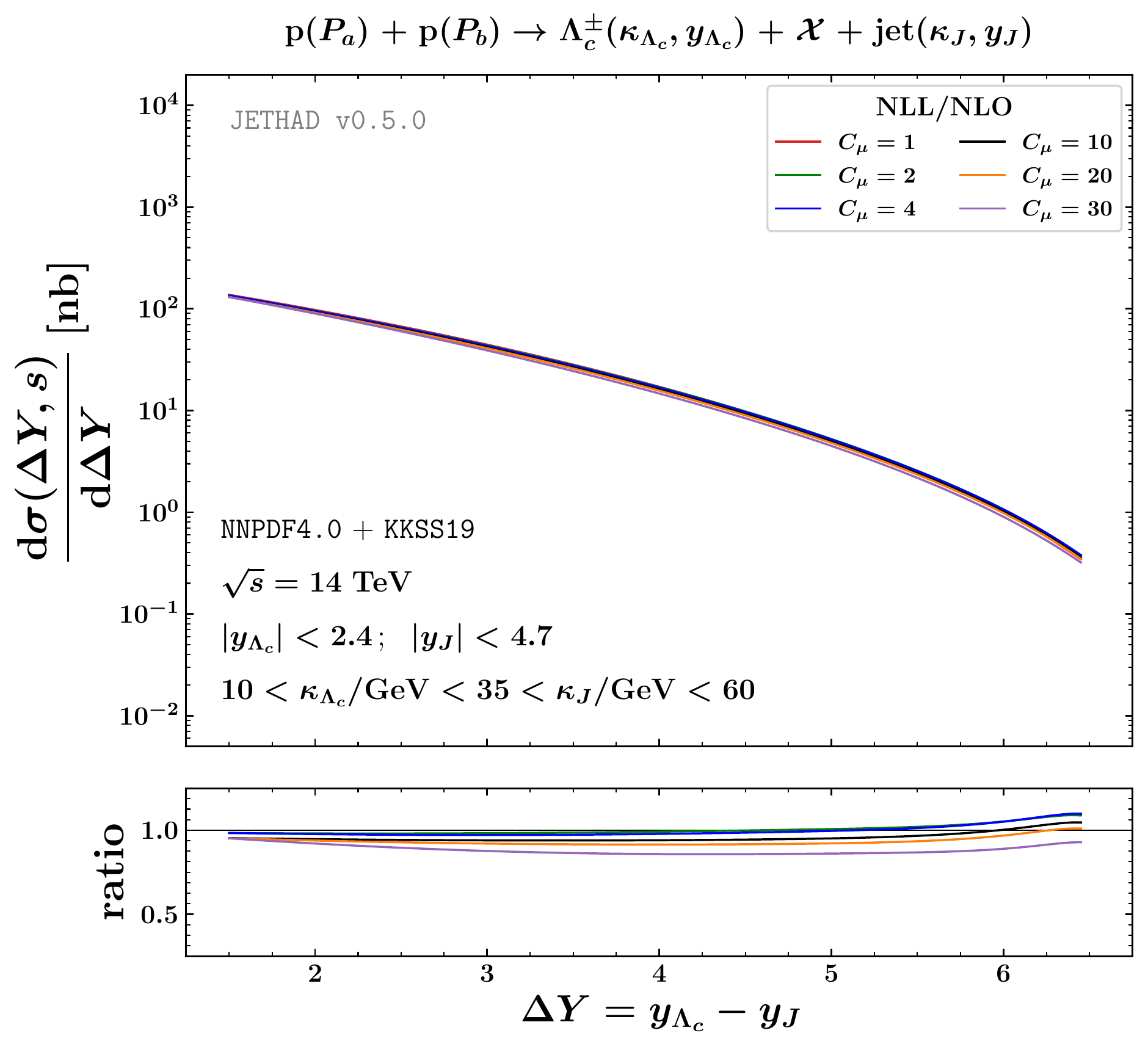}
%   \hspace{0.05cm}
   \includegraphics[scale=0.41,clip]{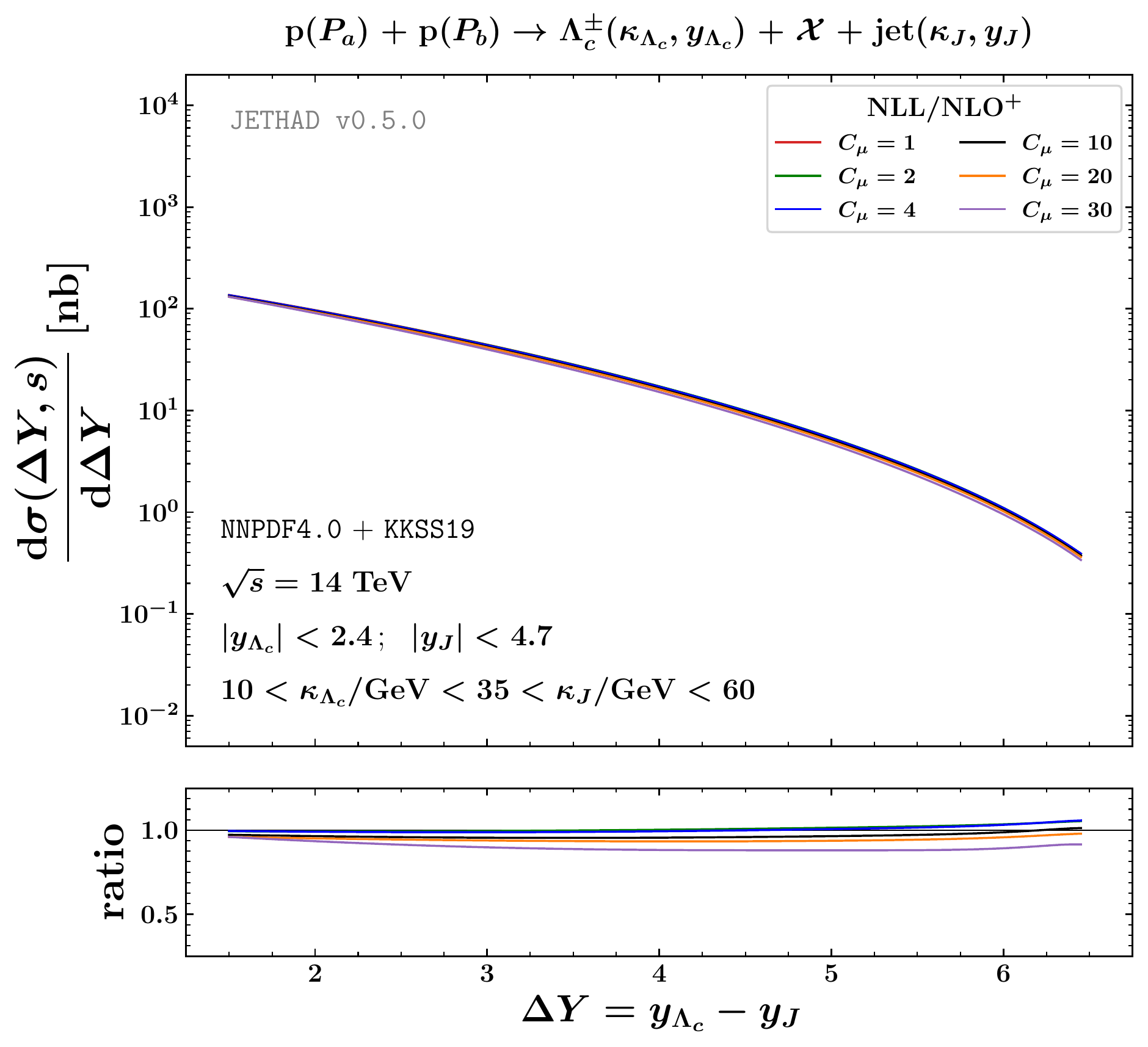}

   \includegraphics[scale=0.41,clip]{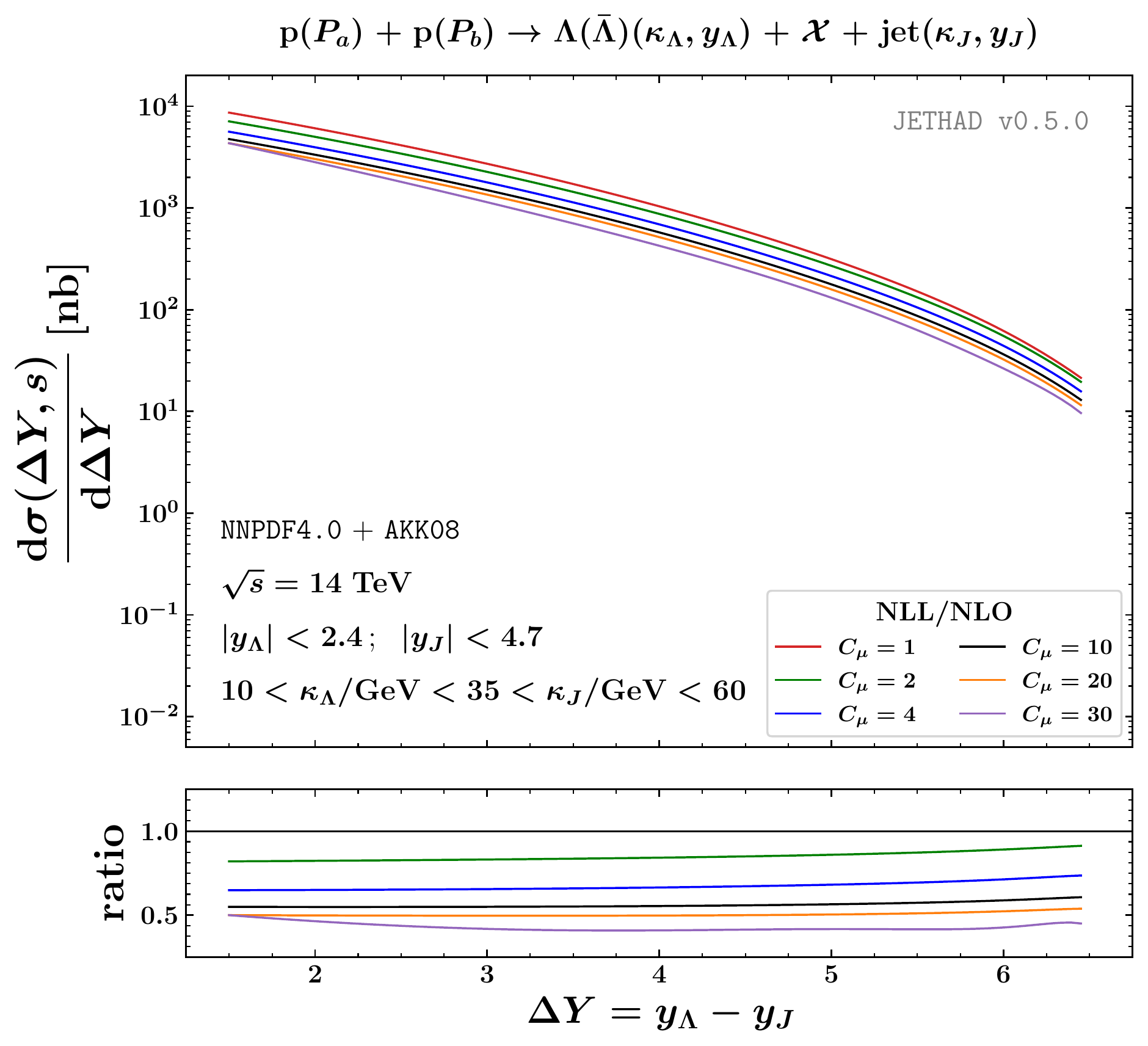}
%   \hspace{0.05cm}
   \includegraphics[scale=0.41,clip]{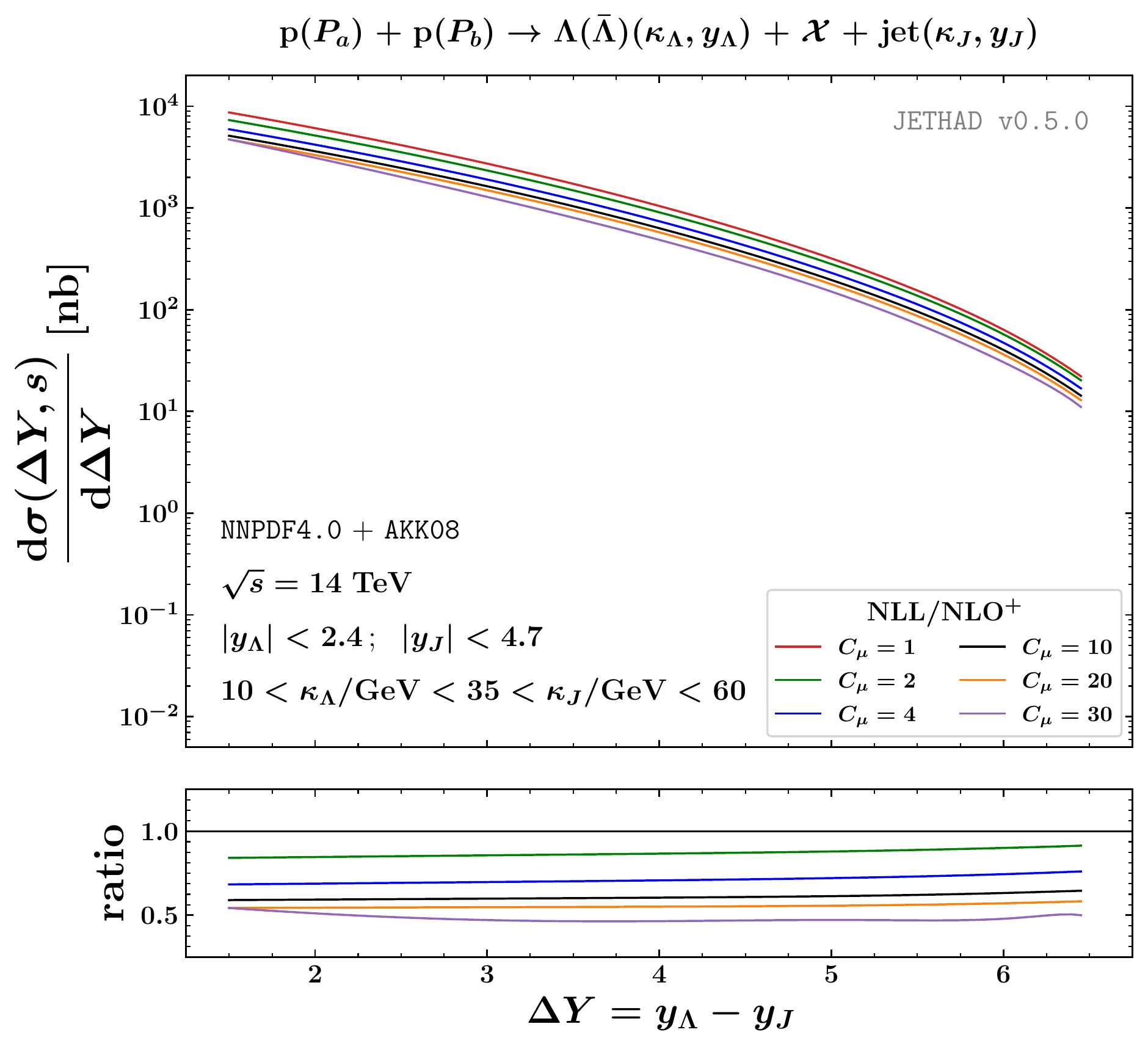}

\caption{$\DY$-dependence of the rapidity distribution within the $\NLL$ (left) and $\NLLp$ (right) accuracy, and for $\sqrt{s} = 14$ TeV. A study on progressive variation of renormalization and factorization scales has been made in the $1 < C_{\mu} < 30$ range for $\Xi$ baryons (upper), $\Lambda_c$ baryons (central), and $\Lambda$ hyperons (lower).}
\label{fig:C0_psv}
\end{figure*}

In upper panels of Fig.\tref{fig:C0} we show the $\DY$-shape of the rapidity distribution for the $\Xi$~plus~jet detection in the kinematic range presented in Section\tref{ssec:kinematics} and for $\sqrt{s} = 14$~TeV.
For the sake of comparison, we present the $\DY$-behavior of the same observable for $\Lambda_c$~plus~jet (central panels) and $\Lambda$~plus~jet (lower panels) production channels.
The downturn at large $\DY$ is a common feature shared by all the distributions, and it rises as a net effect of two competing trends. Indeed, as predicted by BFKL, although high-energy resummed off-shell hard factors strengthen with $\DY$, and thus with energy, their collinear convolution with PDFs and FFs in the impact factors strongly suppresses that upturn.
In left (right) panels we compare $\NLL$ ($\NLLp$) predictions with pure $\LL$ results as well as with corresponding high-energy fixed-order $\HENLO$ ($\HENLOp$) calculations.
Ancillary panels below main plots show reduced $\DY$-distributions, namely cross sections divided by their central value, taken at $C_\mu = 1$. This helps to better visualize the relative size of scale-uncertainty bands.
We observe that NLL bands are uniformly smaller than LL ones.
Furthermore, all bands related with $\Lambda$-hyperon emissions are larger than corresponding ones for $\Lambda_c$- and $\Xi$-baryon detections.
These features indicate that the energy-resummed series gains stability when NLL corrections are accounted for, and that the stabilization mechanism coming from gluon FFs plays a role.
As a general remark, we note that the reached stability is not in the whole range of $\DY$. Indeed, while NLL bands are almost overlapped to LL ones in the low-$\DY$ region, their mutual distance becomes wider and wider as $\DY$ grows. This pattern turns out to be in line with previous analyses on semi-hard heavy-flavor production, where the impressive stability of cross sections on NLL corrections observed in di-hadron production channels (double $\Lambda_c$ baryons,\tcite{Celiberto:2021dzy}, double bottom-flavored hadrons\tcite{Celiberto:2021fdp}, and double vector quarkonia\tcite{Celiberto:2022dyf}) is partially spoiled when a heavy bound state is emitted in association with a jet.
Furthermore, although the discrepancy between $\NLL$ and $\NLLp$ distributions is very small for all the considered final states, it numerically grows with $\DY$, passing from roughly 0.5\% at $\DY \simeq 2$ to almost 5\% at $\DY \simeq 6$, with the $\NLLp$ results constantly staying below the $\NLL$ ones.
This gives us a clue that possible effects coming from Sudakov-type logarithms, enhanced when parton longitudinal fractions become closer and closer to one, are present. These \emph{threshold} logarithms, which are systematically neglected by our hybrid factorization, become relevant in the large-$\DY$ range and they must be resummed to all orders\tcite{Sterman:1986aj,Catani:1989ne,Catani:1996yz,Bonciani:2003nt,deFlorian:2005fzc,Ahrens:2009cxz,deFlorian:2012yg,Forte:2021wxe,Mukherjee:2006uu,Bolzoni:2006ky,Becher:2006nr,Becher:2007ty,Bonvini:2010tp,Ahmed:2014era,Banerjee:2018vvb}.
Combining the resummation of energy and threshold logarithms is not an easy task. While such a double-resummation procedure was set up for Higgs-boson rapidity-inclusive rates\tcite{Bonvini:2018ixe,Ball:2013bra,Mangano:2016jyj}, its extension to two-particle rapidity-differential distributions, as the ones investigated in this article, leads to difficulties rising when the analytic double-counting removal procedure is performed in the Mellin space.
This represents a relevant development to be carried out in more formal, future studies.

For all the considered channels bands for ${\rm HE}\mbox{-}{\rm NLO^{(+)}}$ cross sections are almost overlapped with NLL/NLO$^{+}$ and LL/LO ones ans, in some cases, they stay in between. Thus, at this level a search for a net disengagement between the resummed signal and the fixed-order background still remains unfulfilled (see Section\eref{ssec:phi}).

To further examine the stabilizing effect coming from collinear FFs (see Section\tref{ssec:stability}),
in Fig.\tref{fig:C0_psv} we study the $\DY$-trend of our rapidity distributions under a progressive variation of $\mu_{R,F}$ scales in a wider range, given by $1 < C_\mu < 30$.
Upper, central and lower plots respectively refer to $\Xi$, $\Lambda_c$ and $\Lambda$ plus jet inclusive emissions.
In the same way as in Fig\tref{fig:C0}, ancillary panels below primary plots contain information about the reduced cross sections, \emph{i.e.} divided by the ones taken at $C_\mu = 1$.
Going from bottom to top, we observe that the $\Lambda$ plus jet $\DY$-distribution strongly depends on the scale parameter $C_\mu$. In particular, it decreases as $C_\mu$ grows, up to lose 60\% magnitude when $C_\mu = 30$.
Conversely, the $\Lambda_c$ plus jet $\DY$-distribution is quite stable on $C_\mu$ variation, its magnitude loss staying from 5\% to 10\% only.
The pattern of the $\Xi$ plus jet $\DY$-distribution stays in between the previous two, namely its magnitude loss does not exceed 35\%.
The founds trend are in line with the statement that the behavior of collinear FFs and, in particular, of the gluon one, determine if and to which level the stabilizing effect is present.
No significant variation of the stabilization pattern is spotted when passing from the $\NLL$ (left panels) to the $\NLLp$ (right panels) representation.

%-----------------------------------------
\subsection{Azimuthal distributions}
\label{ssec:phi}
%-----------------------------------------

\begin{figure*}[!p]
\centering

\begin{multicols}{2}

 \includegraphics[scale=0.41,clip]{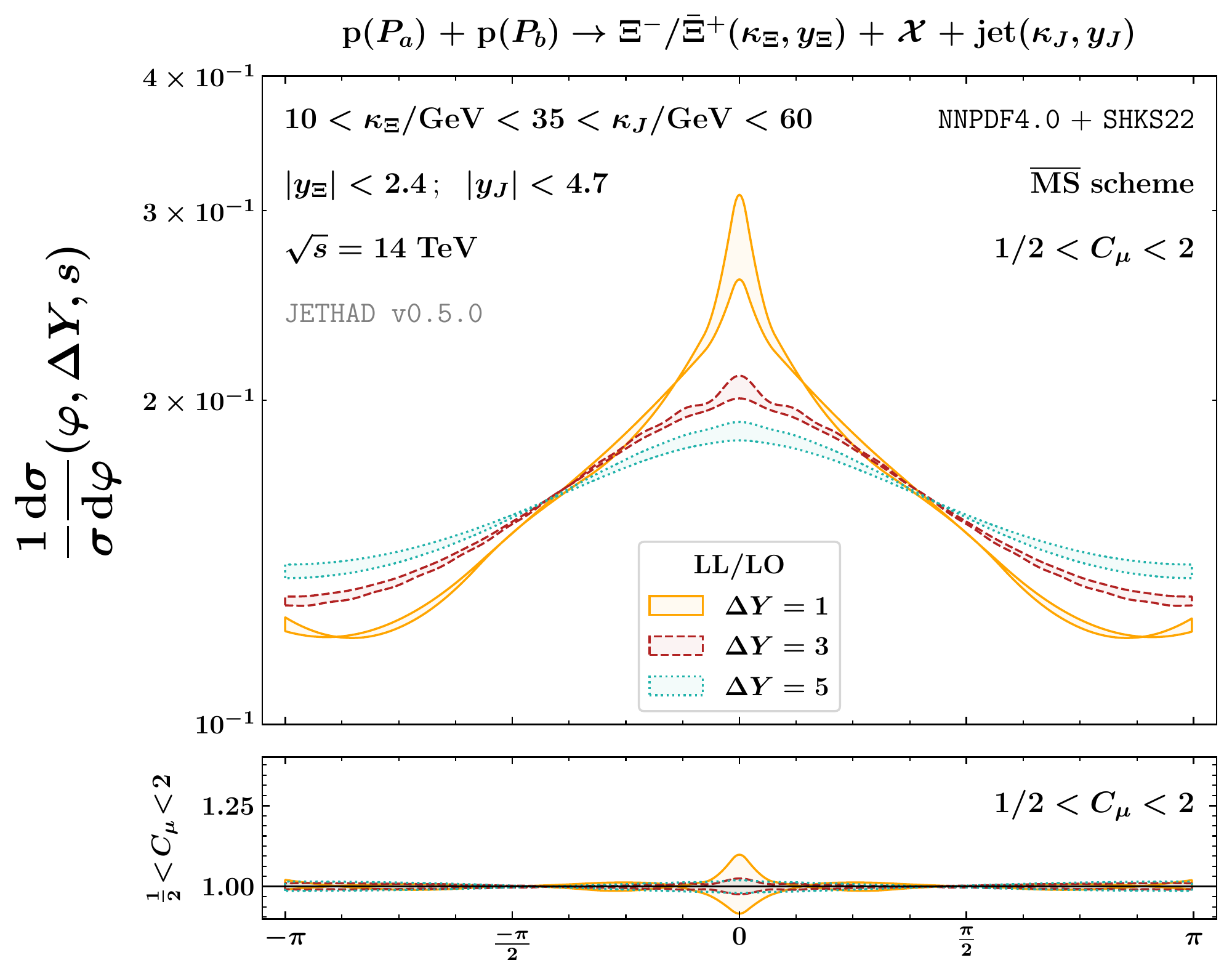}

 \includegraphics[scale=0.41,clip]{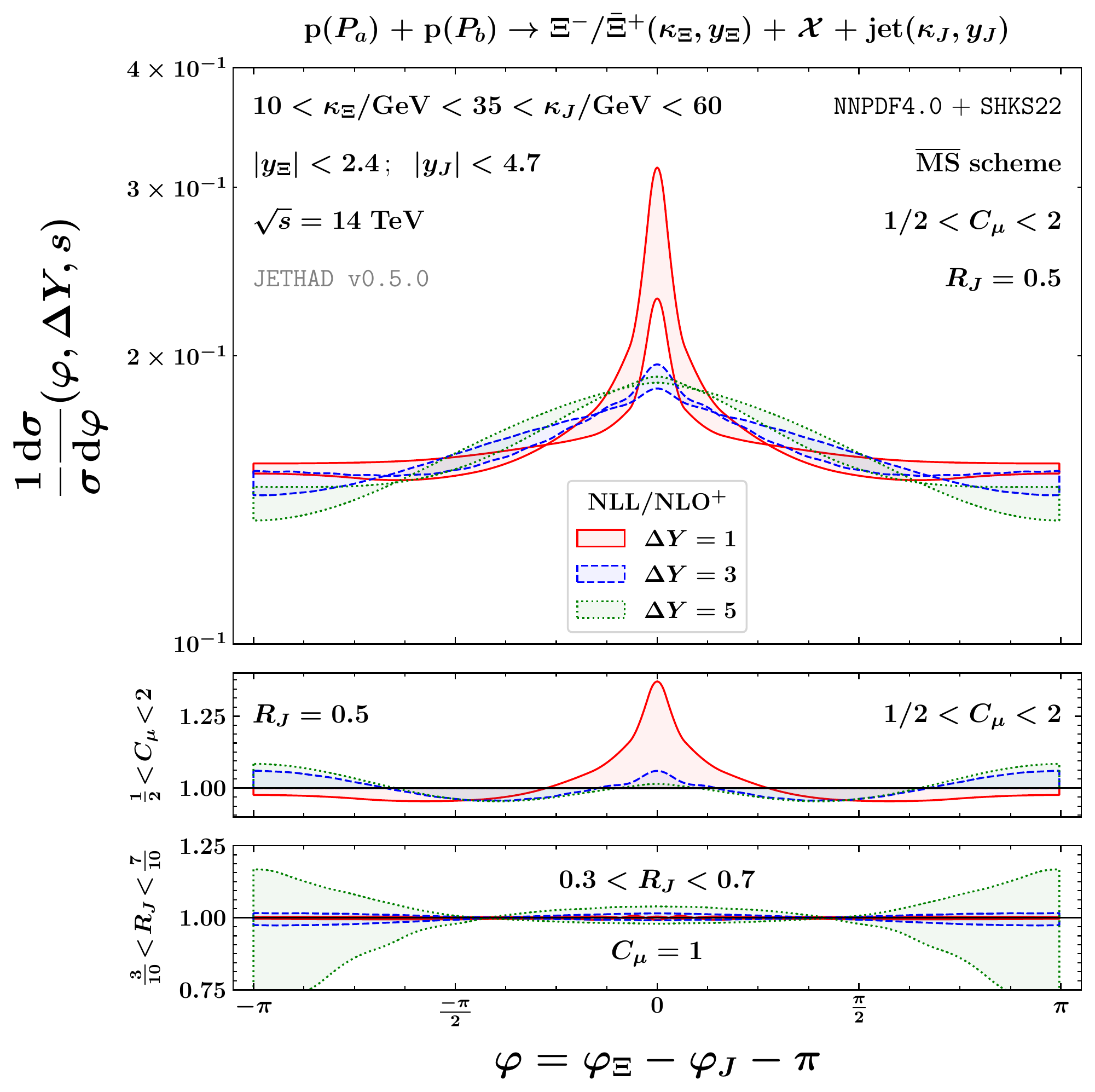}

\end{multicols}

\caption{Azimuthal distribution for the $\Xi^-/\bar\Xi^+$~plus~jet detection at $\DY=1,3,5$, for $\sqrt{s} = 14$ TeV, and within the $\LL$ (left) and $\NLLp$ (right) accuracy.
Shaded bands embody the combined effect of renormalization- and factorization-scale variation in the $1 < C_{\mu} < 2$ range and of phase-space numerical multidimensional integration.}
\label{fig:phi}
\end{figure*}

In Fig.\tref{fig:phi} we show the behavior of the azimuthal distributions at $\DY=1, 3, 5$. The first ancillary panels right below main plots exhibit reduced $\varphi$-distributions, \emph{i.e.} azimuthal cross sections divided by their central value, taken at $C_\mu = 1$.
For the sake of simplicity, we consider the $\NLLp$ representation only.
Form the inspection of our plots, we fairly observe the emergence of high-energy dynamics.
All $\varphi$-distributions exhibit a peak at $\varphi = 0$, namely when the $\Xi$ baryon and the jet are emitted in back-to-back configurations.
The peak height shrinks as $\DY$ increases, and the distribution width broadens. This a consequence of the onset the BFKL dynamics. Indeed, when $\DY$ grows, the weight of gluons strongly ordered in rapidity, predicted by the resummation, increases. 
This reduces the azimuthal-angle correlation between the baryon and the jet, so that the number of back-to-back events diminishes.
We note that the discrepancy among results taken at different values of $\DY$ is larger in the LL case with respect to the NLL one. This is in line with previous findings in the context of semi-hard reactions involving hadron emissions (see, \emph{e.g.}, Refs.\tcite{Celiberto:2017ptm,Bolognino:2018oth,Celiberto:2020wpk,Celiberto:2022dyf}), where a recorrelation effect due to genuine NLL contributions was observed.
Besides the lowering-with-$\DY$ trend of the peak, which is a common feature of all the semi-hard final states investigated so far, our novel $\Xi$~plus~jet detection process exhibit some peculiar features. Indeed, while $\Xi$-particle collinear FFs lead to a stabilization pattern, typical of heavy-flavored hadron species, the pattern of $\varphi$-distributions sensitive to $\Xi$ emissions is more similar to the one typical of light-flavored objects.
It is easy to see that distributions of Fig.\tref{fig:phi} are more similar to corresponding ones for Mueller--Navelet jet and light-hadron detections\tcite{Celiberto:2020wpk}, with milder peaks and wider widths than the ones observed in vector-quarkonium\tcite{Celiberto:2022dyf} and $B_c^{(*)}$-meson\tcite{Celiberto:2022keu} hadroproductions.
The duality of phenomenological aspects emerging in $\Xi$~plus~jet studies, leading both to stabilizing features typical of heavy-flavor emissions and to distribution patterns close to light-flavor detections, makes our process novel and intriguing. Further investigations on the origin and interplay of these aspects will help us to deepen our understanding of high-energy QCD.

Another intriguing aspect is the $(\ln R_J)$-dependence of our observables. Although, from a general QCD viewpoint, it is natural to set $R_J \sim {\cal O}(1)$ (see, \emph{e.g.}, Ref.\tcite{Catani:1993hr}), smaller values are adopted in most practical applications in hadron scatterings. Indeed, small-$R_J$ choices are helpful to dampen the so-called \emph{pileup} contaminations rising from multiple hadron-hadron interactions, as well as to resolve the jet substructure.
The small-$R_J$ limit leads to the enhancement of terms proportional to $(\alpha_s \ln 1/R_J^2)^n$, which need to be resummed to all orders\tcite{Dasgupta:2014yra,Dasgupta:2016bnd,Cacciari:2015vwj,Banfi:2012jm,Banfi:2015pju,Kang:2016mcy,Liu:2017pbb}.
The question whether $(\ln 1/R_J^2)$-contributions are important for our high-energy observables becomes particularly relevant for angular distributions. Indeed, any variation of the jet-radius size can have an influence on the number of back-to-back events.
With our choice, $R_J = 0.5$, witch matches the CMS experimental setup\tcite{Khachatryan:2016udy}, and for a typical value
of the running coupling, say $\alpha_s \simeq 0.25$, one has $\alpha_s \ln 1/R_J^2 \simeq 0.35$, which is not so large.
Therefore, by relying on our formalism where these logarithms are encoded in the jet impact factor but they are not resummed, we can estimate the effect of varying the jet radius on a reference range, say $0.3 < R_J < 0.7$.
The second ancillary panel below the right plot of Fig.\tref{fig:phi} shows how varying $R_J$ reflects on the $\NLLp$ reduced $\varphi$-distribution taken at natural scales, $C_\mu = 1$.
Our angular observable increases with $R_J$, but the effect is generally small, say below $3\%$. The only exception is represented by the case corresponding to the largest value of the rapidity interval, $\DY = 5$. Here, the uncertainty band generated by varying $R_J$ becomes larger and larger when $\varphi \gtrsim \pi/2$, up to roughly $30\%$. While further studies are needed to more deeply investigate this feature, a possible explanation can be already guessed. We note that $(\ln 1/R_J^2)$-terms enter the NLO forward-jet impact-factor correction as multiplicative factors for $\zeta^{-2\gamma} \equiv \zeta^{1-2i\nu}$ and for all the $P_{ij}$ splitting functions (see Eq.~(36) of Ref.~\cite{Caporale:2012ih}).
Since $P_{ij}$ kernels do not depend on the angular kinematics, the origin of the increased sensitivity on $\ln 1/R_J^2$ with $\DY$ must be sought in the integration over $\zeta$. Indeed, the lower bound of that integral is $x_J$, whose value faster enters the already-mentioned threshold region for large $\DY$-values.
Therefore, variations of $R_J$ could magnify the instabilities already present in our formalism due to nonresummed threshold logarithms.
Future analyses will focus on encoding the jet-radius resummation into our high-energy formalism, as well as to explore possible common ground with studies on jet angularities\tcite{Luisoni:2015xha,Caletti:2021oor,Reichelt:2021svh}.

A key aspect emerging from the previous discussion is making use of azimuthal distributions as useful tools to access the intersection corners among different resummation approaches.
More in particular, $\varphi$-differential observables gives us a direct access to almost-back-to back configurations generating the peak region of Fig.\tref{fig:phi}. Here, due to collinear enhancement, final-state soft gluons tend be emitted aligned to jet directions. A large part of this radiation falls into jet cones, thus becoming part of the jets. The remaining soft-gluon radiation standing lightly outside jet cone generates positive angular asymmetries which can have a sizable impact on azimuthal observables\tcite{Hatta:2021jcd,Hatta:2020bgy}.
Its net effect is the rise of double and single Sudakov-type logarithms in the small transverse-momentum imbalance between the two detected objects. Our hybrid factorization does not embody the resummation of these logarithms.
Since in our studies a large rapidity interval is required, so that each of the two final-state particles stems from a distinct fragmentation region, soft-gluon logarithms appear only at the cross-section level and not at the impact-factor one. Therefore, accounting them for is not straightforward.
Conversely, the semi-inclusive emission of a hadronic system made of two forward particles represents a more favorable configuration. In that case, the Sudakov resummation of small transverse-momentum imbalances can be performed directly inside the impact factor. 
Advancements in this direction have been made in the context of inclusive dijet  tags in deeply inelastic electron-nucleus scatterings via the saturation framework (see Ref.\tcite{Wallon:2023asa} for a recent overview) and they can be also planned for diffractive dihadron detections\tcite{Fucilla:2022szm,Fucilla:2022wcg}.

%-----------------------------------------
\subsection{Transverse-momentum distributions}
\label{ssec:pT}
%-----------------------------------------

\begin{figure*}[!t]
\centering

   \includegraphics[scale=0.41,clip]{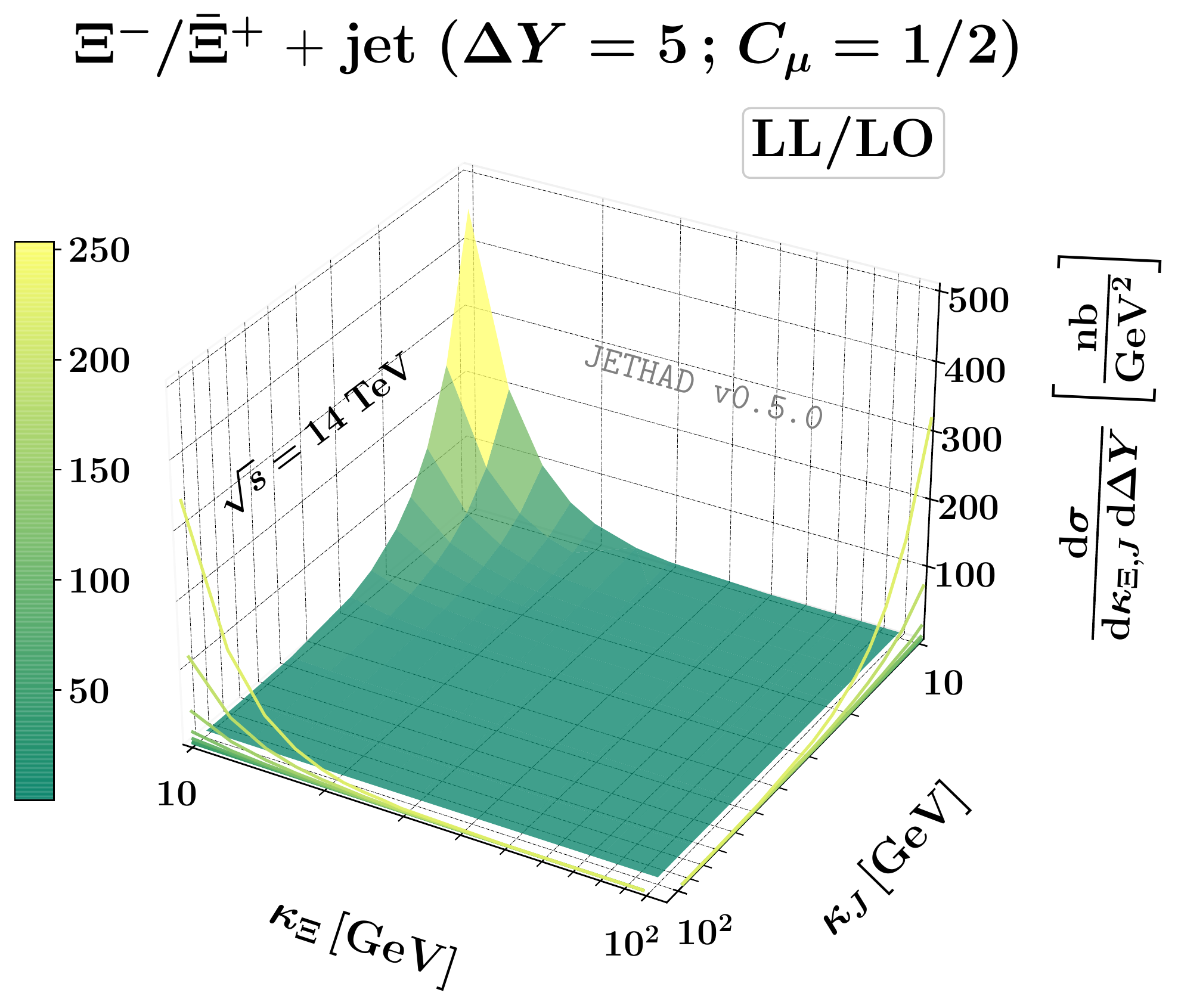}
   \hspace{0.25cm}
   \includegraphics[scale=0.41,clip]{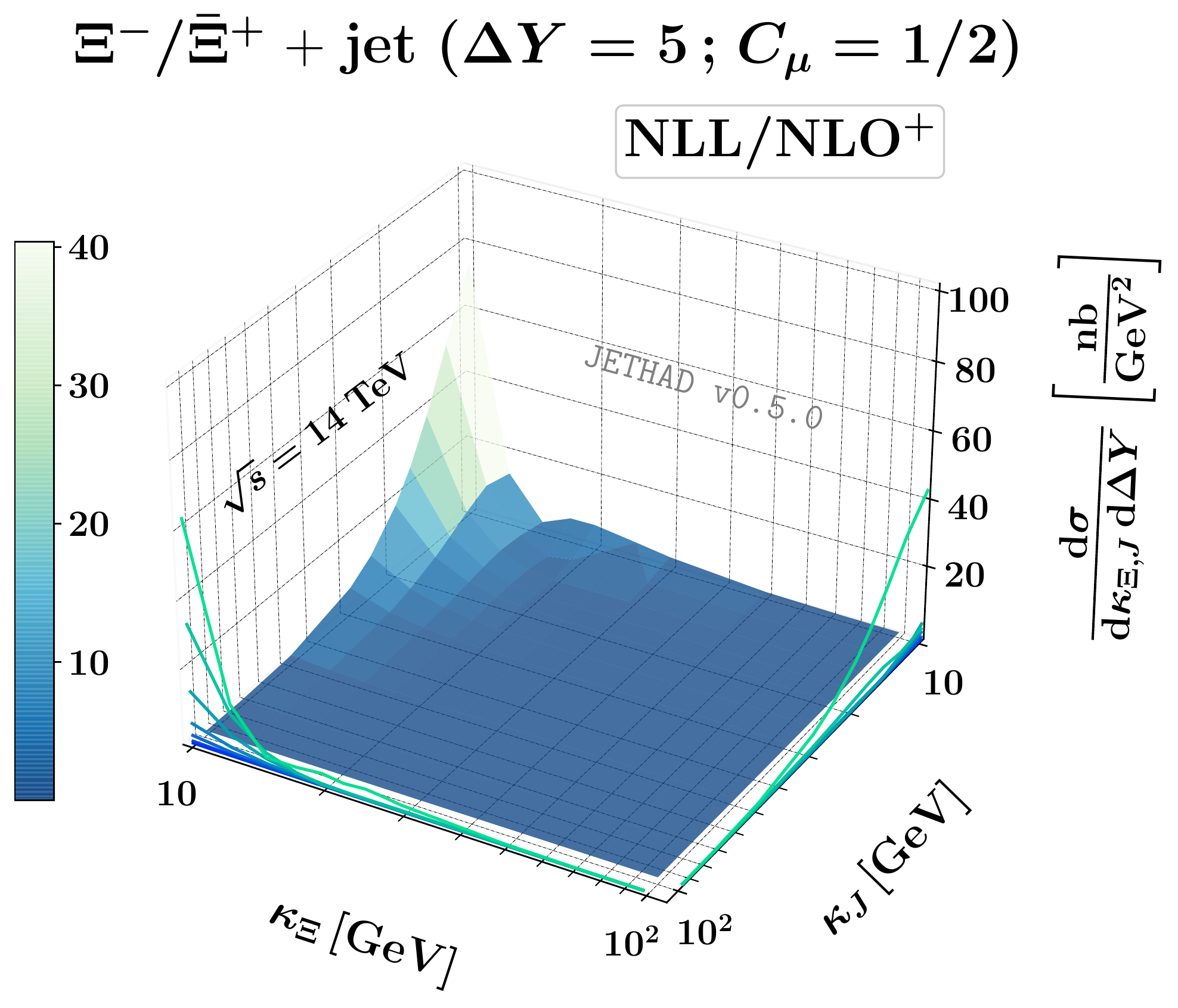}
   \vspace{0.15cm}

   \includegraphics[scale=0.41,clip]{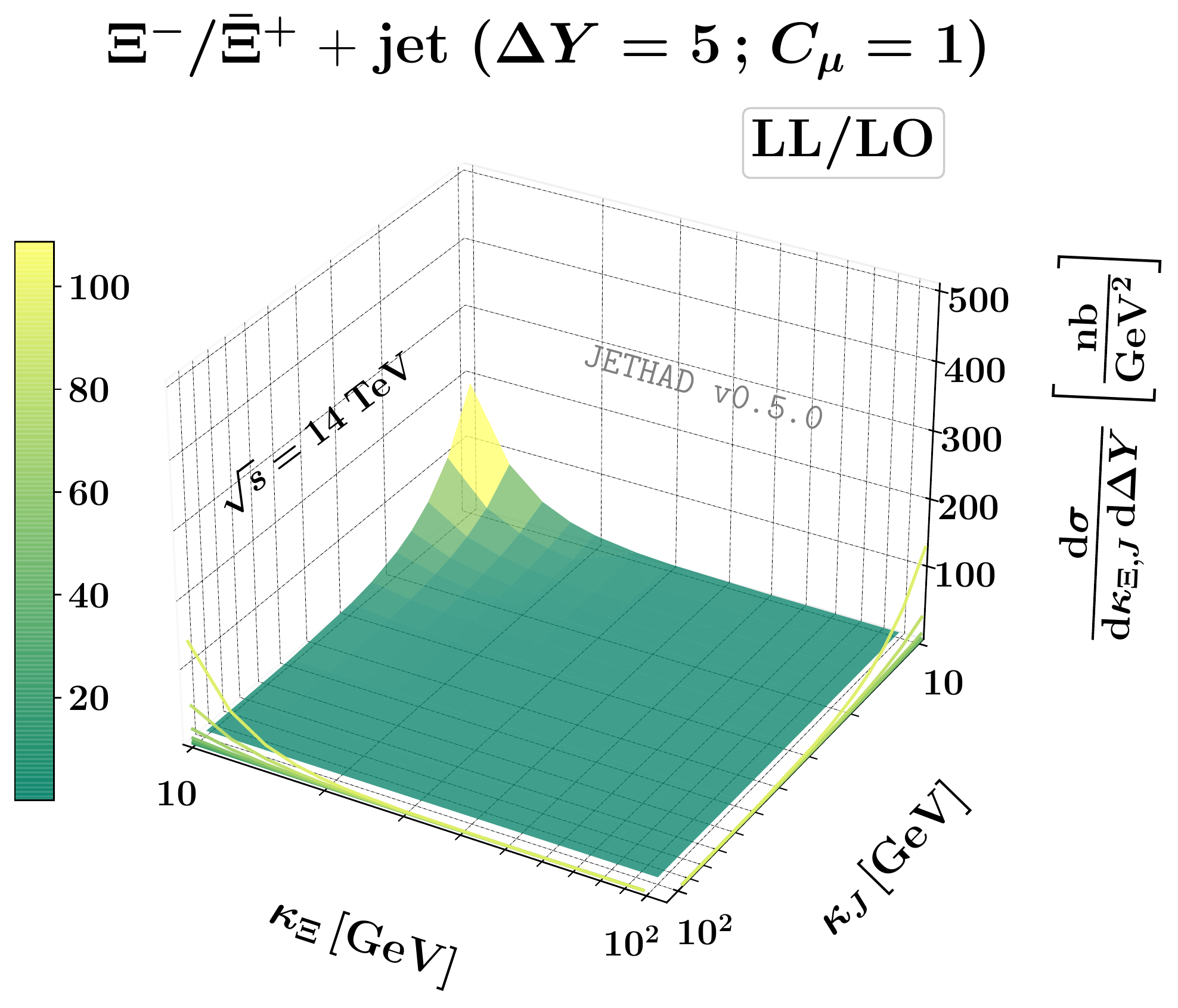}
   \hspace{0.25cm}
   \includegraphics[scale=0.41,clip]{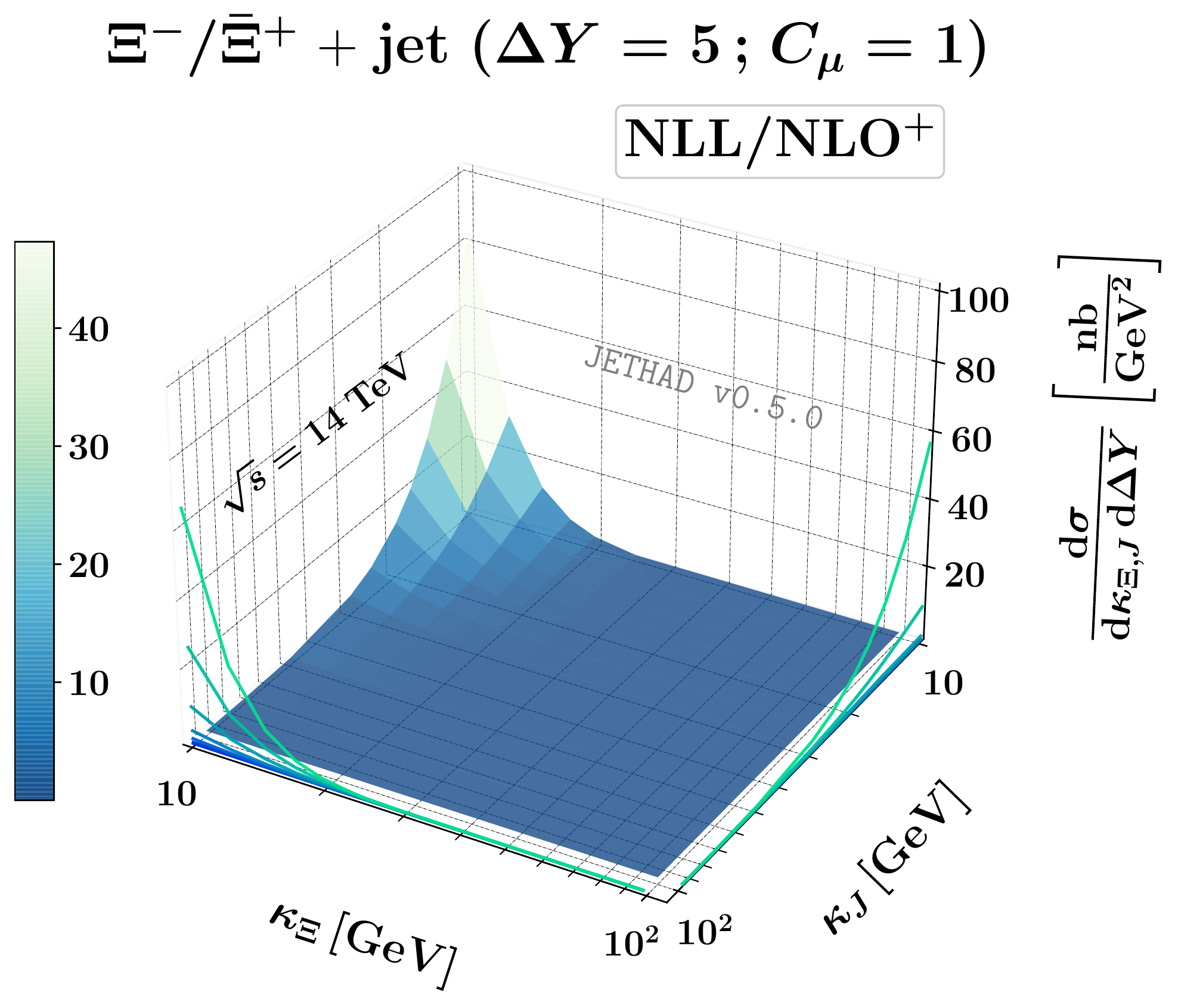}
   \vspace{0.15cm}

   \includegraphics[scale=0.41,clip]{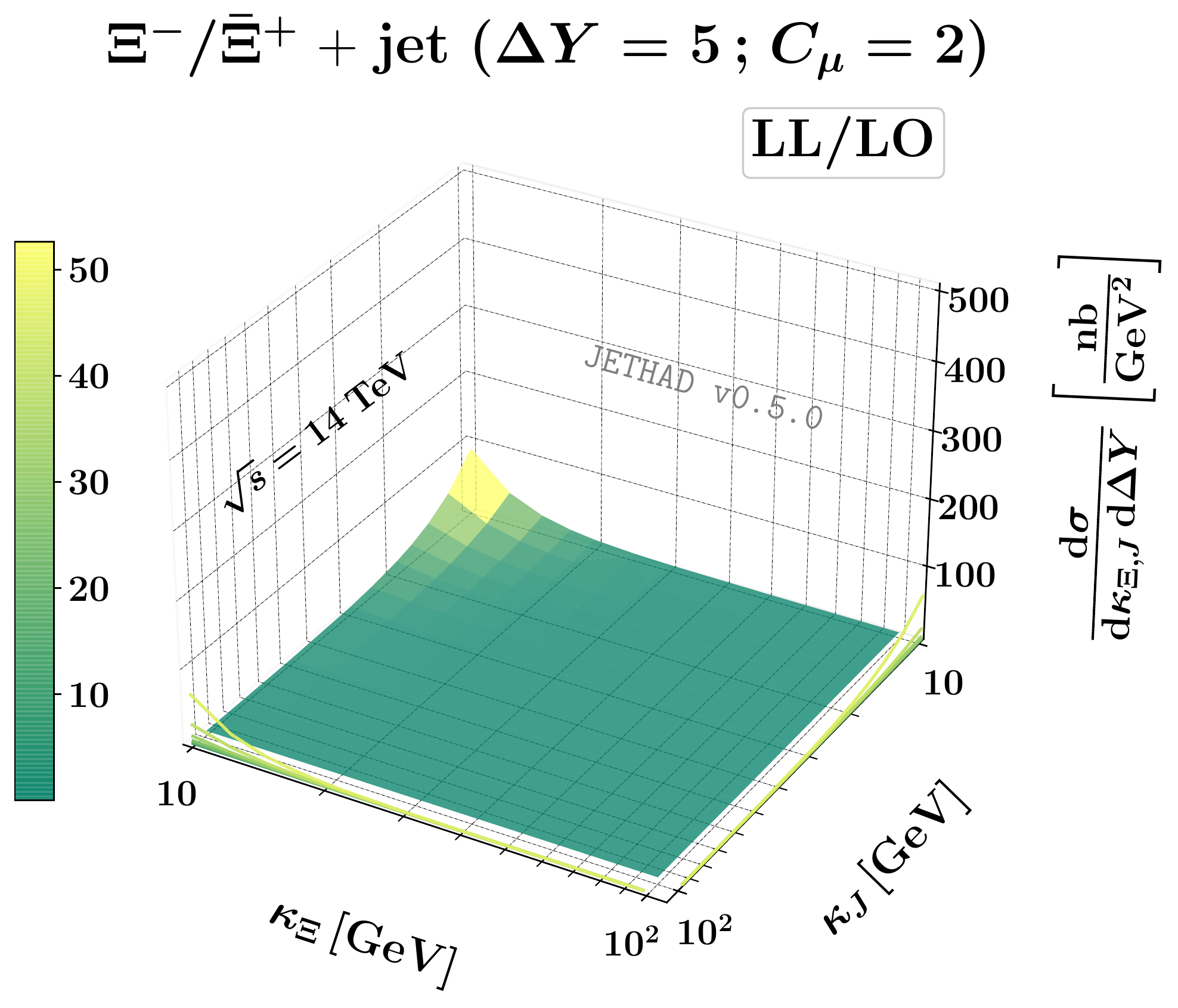}
   \hspace{0.25cm}
   \includegraphics[scale=0.41,clip]{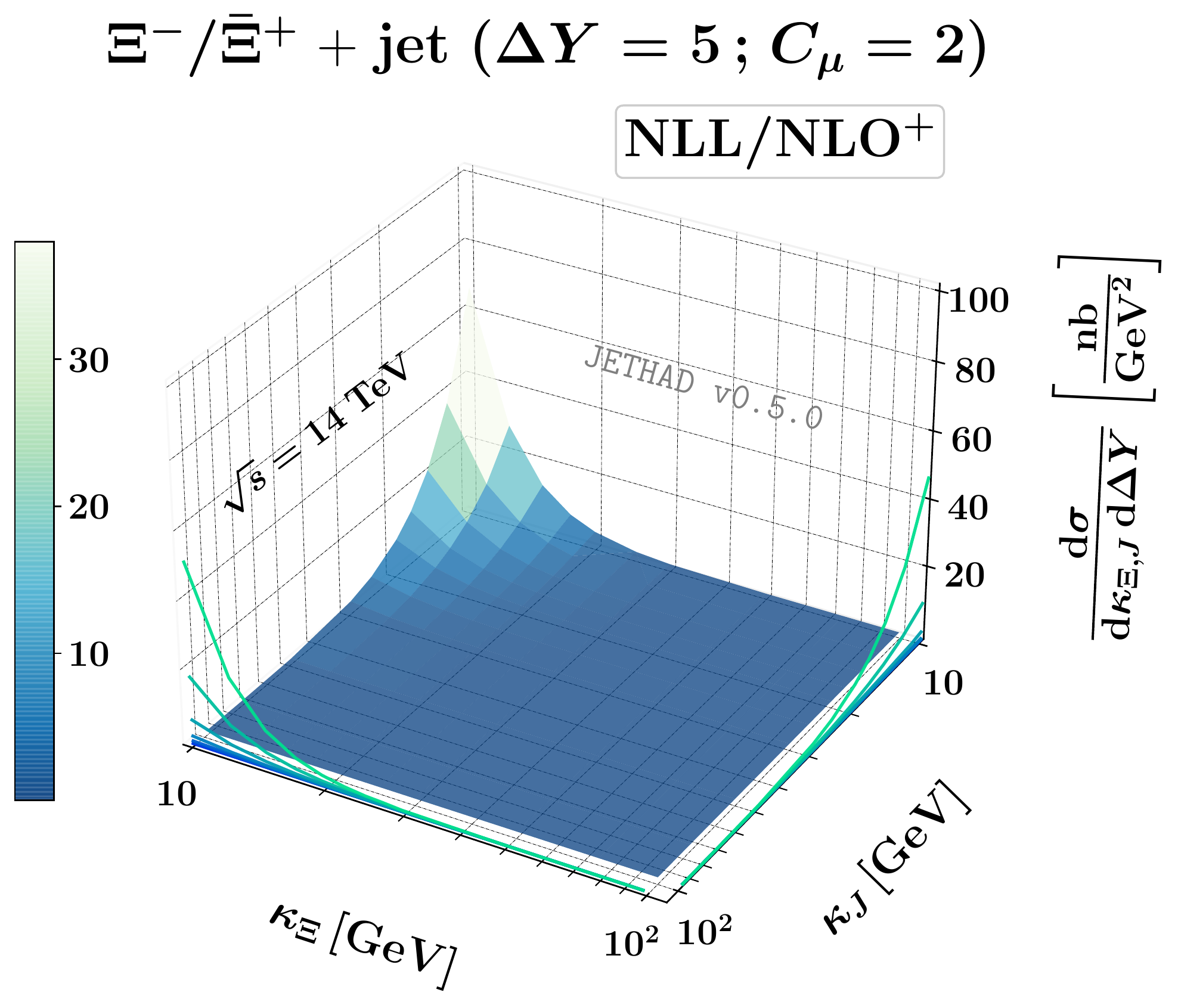}
   \vspace{0.15cm}

\caption{Double differential transverse-momentum distribution for the $\Xi^-/\bar\Xi^+$~plus~jet detection at $\DY=5$, $\sqrt{s} = 14$ TeV, within the $\LL$ (left) and $\NLLp$ (right) accuracy. The $C_\mu$ energy-scale parameter ranges from 1/2 to 2 (from top to bottom).}
\label{fig:Y5-2pT0}
\end{figure*}

\begin{figure*}[!t]
\centering

   \includegraphics[scale=0.41,clip]{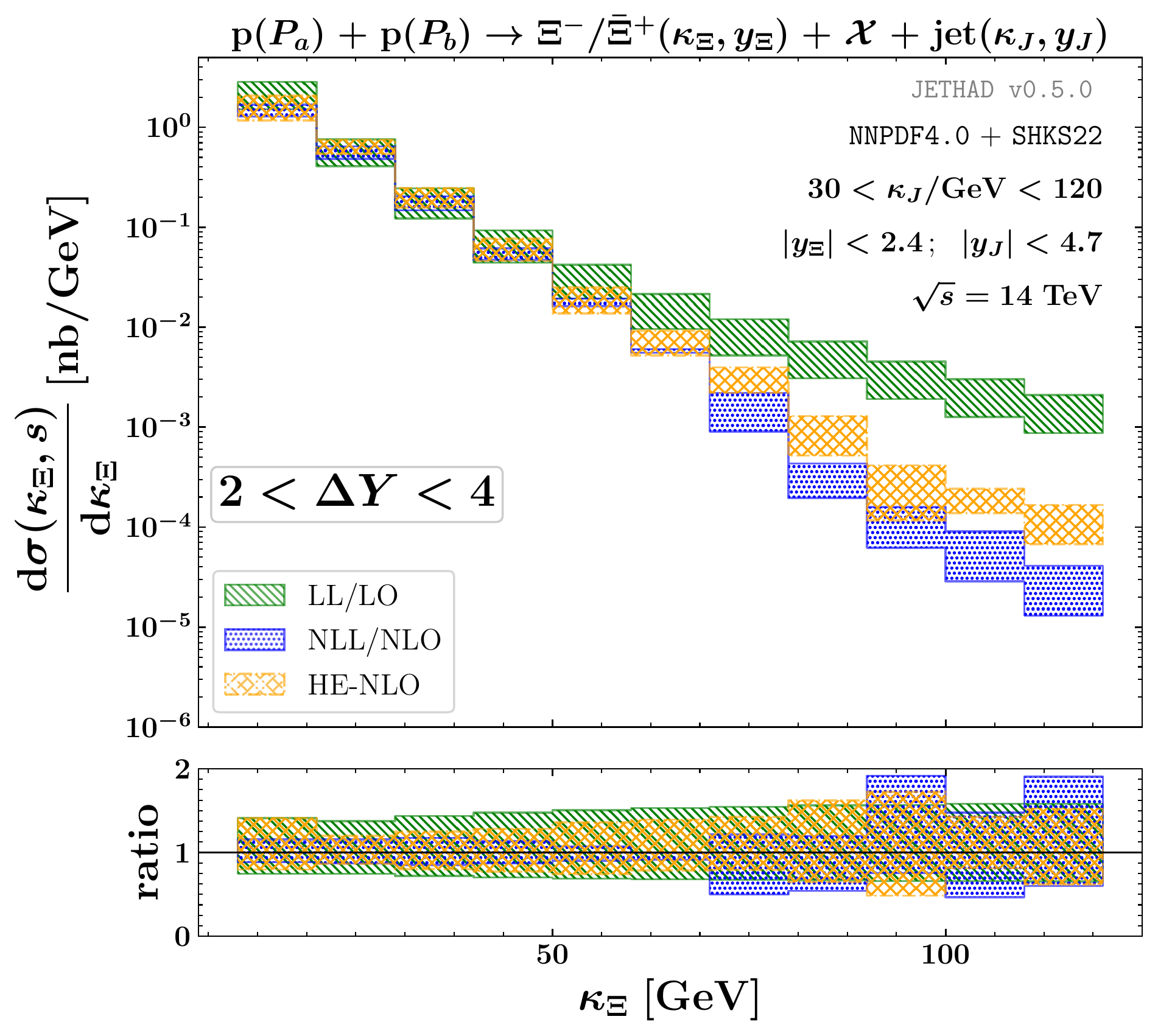}
   \hspace{0.25cm}
   \includegraphics[scale=0.41,clip]{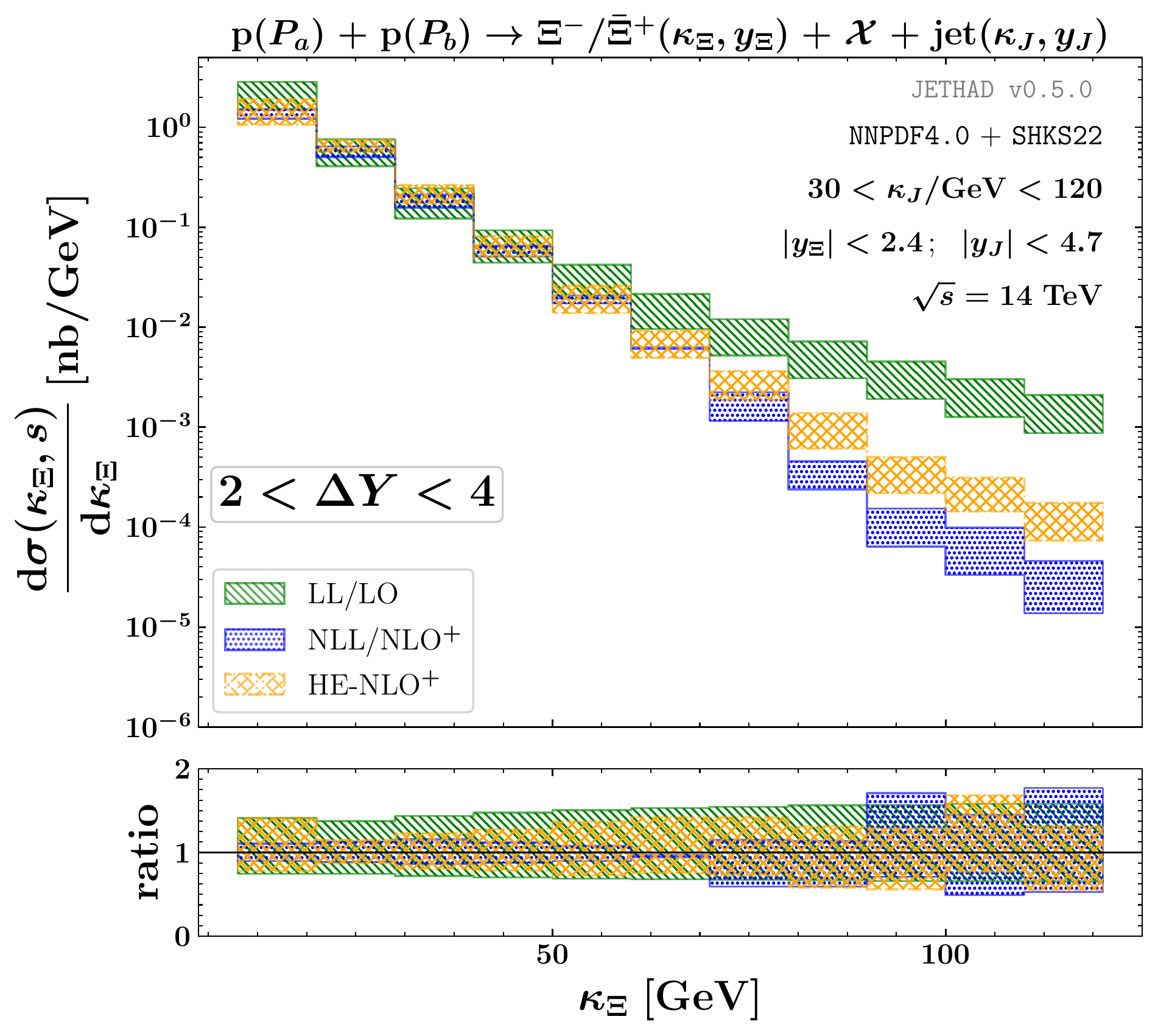}
   \vspace{0.15cm}

   \includegraphics[scale=0.41,clip]{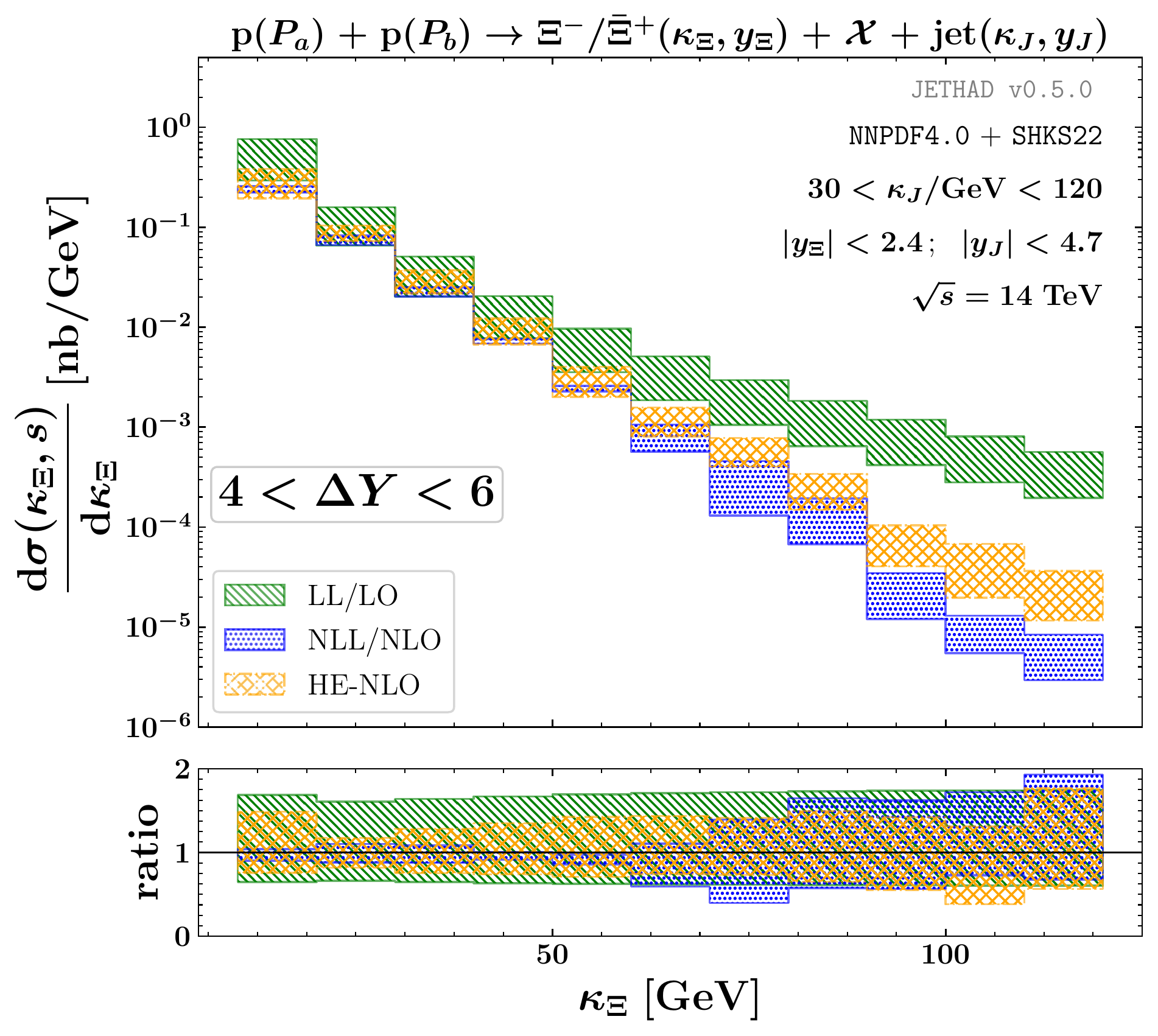}
   \hspace{0.25cm}
   \includegraphics[scale=0.41,clip]{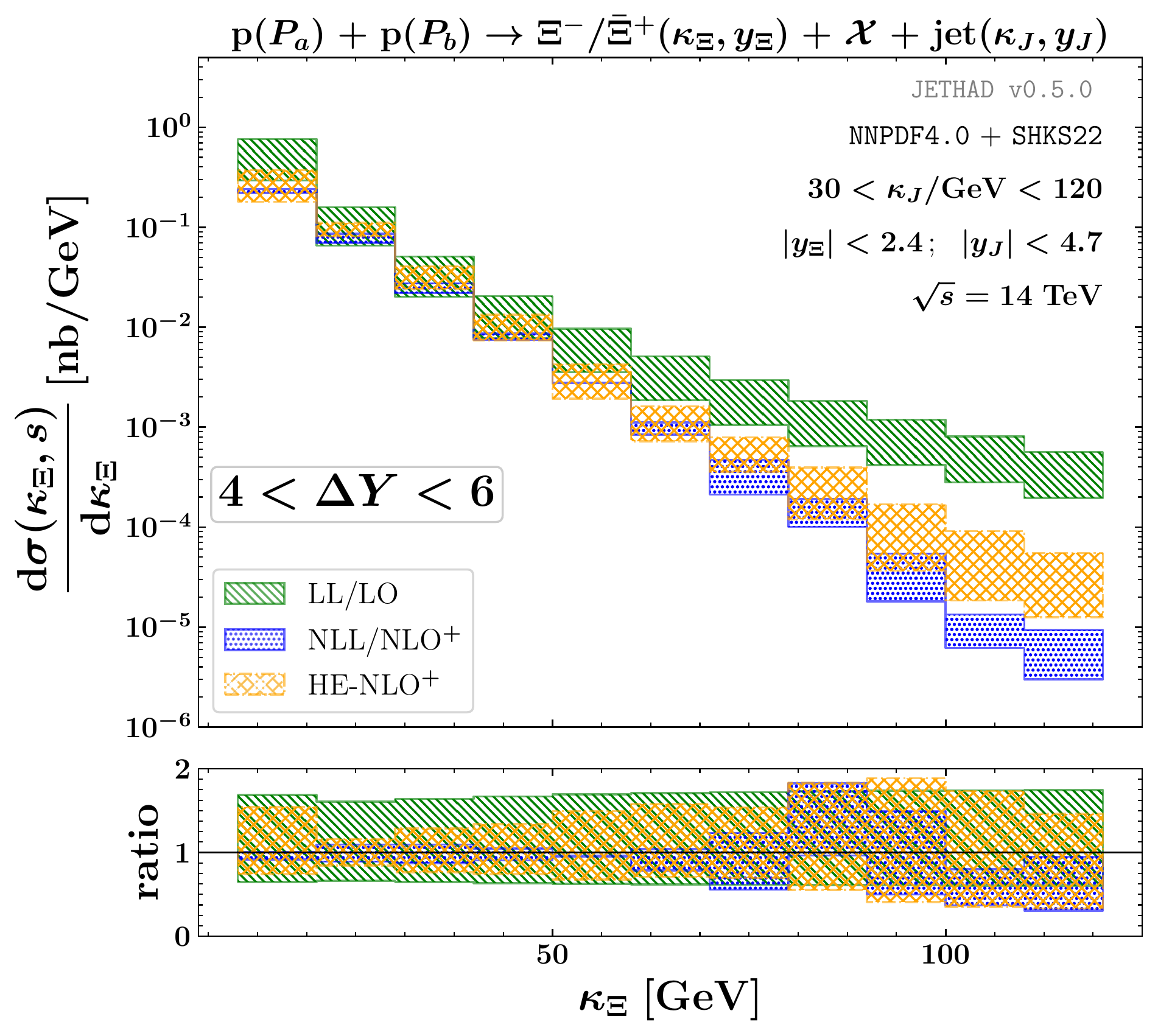}   \vspace{0.15cm}

\caption{Behavior of the $|\vec \kappa_\Xi|$-differential distribution at $\sqrt{s} = 14$ TeV and within the $\NLL$ (left) and $\NLLp$ (right) accuracy, for $2 < \DY < 4$ (upper) and $4 < \DY < 6$ (lower). Shaded bands embody the combined effect of renormalization- and factorization-scale variation in the $1 < C_{\mu} < 2$ range and of phase-space numerical multidimensional integration.}
\label{fig:pT_Xi}
\end{figure*}

In Fig.\tref{fig:Y5-2pT0} we present
predictions for the double differential transverse-momentum cross sections at $\DY=5$. For the sake of simplicity, we consider the $\NLLp$ representation only. 
Our distributions exhibit a very fast decreasing behavior when the two transverse momenta, $|\vec \kappa_\Xi|$ and $|\vec \kappa_J|$, grow or when their mutual distance increases. LL results (left panels) are much more sensitive to scale variations than corresponding ones for $\Delta Y$- and $\varphi$-distributions.
Indeed, they globally decrease with $C_\mu$ (from upper to lower panels).
Conversely, NLL predictions (right panels) tend to oscillate around $C_\mu = 1$, which seems to act as a critical point. This represents a clear indication that a stability on scale variations of our double $\kappa_T$-observables distributions is reached when energy logarithmic corrections are taken at NLL.
In all cases, we observe the absence of any peak. However, a peak could be present in the very small transverse-momentum range, \emph{i.e.} in the region dominated by TM-resummation effects, which has been excluded from our study.
Surface 3D plots are complemented by 2D contour projections showing the behavior of our distributions at $|\vec \kappa_\Xi| = 0$ and at $|\vec \kappa_J| = 0$.
The information gathered from the inspection on these projections at low/intermediate $\kappa_T$ is that cross sections are smaller when $|\vec \kappa_\Xi| < |\vec \kappa_J|$ than when $|\vec \kappa_\Xi| > |\vec \kappa_J|$.
This reflect the fact that cross sections are generally larger when a light hadron is produced rather than a jet (see, \emph{e.g}, Refs.\tcite{Bolognino:2018oth,Celiberto:2020wpk}). As pointed out in the context of bottom-flavored hadrons plus jets\tcite{Celiberto:2021fdp}, this hierarchy of predictions eventually reverts when the transverse momentum increases.

In Fig.\tref{fig:pT_Xi} we show the $|\vec \kappa_\Xi|$-differential cross section integrated in two ranges for the rapidity interval, $2 < \DY < 4$ (upper panels) and $4 < \DY < 6$ (lower panels).
In left (right) panels, $\NLL$ ($\NLLp$) results are compared with corresponding $\LL$ and $\HENLO$ ($\HENLOp$) calculations.
To propose realistic distribution configurations that can be easily compared with future experimental data to be collected at the LHC and its high-luminosity upgrade, we consider $|\vec \kappa_\Xi|$ bins with a fixed length of 10~GeV and in the range from 10 to 120~GeV.
The information gathered from these plots significantly extends and complements the one encoded in our double differential cross sections of Fig.\tref{fig:Y5-2pT0}.

Two subregions can be distinguished.
The first one is the low-to-moderate $|\vec \kappa_\Xi|$ region, which ranges from 10 to roughly 60 GeV.
Here, NLL bands are almost everywhere nested inside LL ones, their size being always much smaller than the LL one, see ancillary panels below primary figures.
This further confirms the impressive stability of the hybrid factorization under higher-order corrections and scale variations.
The found pattern is expected. 
Indeed, although the jet transverse momentum is integrated in a larger range,  30~GeV~$< |\vec \kappa_J| <$~120~GeV, its main contribution to cross section comes from the lower spectrum of values. 
A similar argument applies for $\kappa_\Xi$. 
Thus, in the considered low-to-moderate subregion the weight of (almost) back-to-back events, generally well described by the BFKL resummation, is large. 
At the same time, since (almost) symmetric transverse-momentum windows are not well suited to disentangle the BFKL pattern from fixed-order contaminations, in this region NLL bands are also overlapped with high-energy NLO ones.

The second region is the moderate-to-large $|\vec \kappa_\Xi|$ one, 60~GeV~$\lesssim |\vec \kappa_\Xi| <$~120~GeV.
Here, the NLL signal decouples from the LL one, thus indicating a potential loss of stability of the hybrid factorization. Indeed, although the number of (almost) back-to back events equals the one typical of the previous region, their weight is smaller. 
On the one hand, this translates in a growth of relevance of asymmetric transverse-momentum configurations, suited to disentangle BFKL from fixed-order results.
On the other hand, kinematic configurations featuring a large mutual distance between $|\vec \kappa_\Xi|$ and $|\vec \kappa_J|$ lead to rising DGLAP-type logarithms as well as \emph{threshold} effects, which are not accounted for in our formalism and could spoil the convergence of the resummed series.
In contrast to the mentioned potential issues, the outcome emerging from the inspection of our distribution is favorable.
Indeed, although being larger than the one in the low-to-moderate region, the width of NLL uncertainty bands encoding the effect of scale variations remains steady.
This corroborates the statement that the hybrid factorization is still valid.
Moreover, the discrepancy between NLL results and high-energy fixed-order computations progressively widens as $|\vec \kappa_\Xi|$ increases, up to make the corresponding uncertainty bands not anymore nested to each other.
Thus, a clear indication comes out that the moderate-to-large $|\vec \kappa_\Xi|$ range is a fertile ground where to hunt for distinctive high-energy imprints on top of the fixed-order background.
As a final remark, we note that, for large $|\vec \kappa_\Xi|$ values, a slight or moderate difference between $\NLL$ and $\NLLp$ predictions becomes more and more evident, as highlighted in the ancillary panels.
Further studies will gauge the impact of this effect in larger transverse-momentum ranges or/and in other transverse-momentum related observables.

%-----------------------------------------
\section{Paving the way toward precision}
\label{sec:conclusions}
%-----------------------------------------

We have proposed the inclusive detection at the LHC of a cascade $\Xi^-/\bar\Xi^+$ baryon in association with a jet, as a new probe channel for the high-energy spectrum of QCD. Their large separation in rapidity and their high transverse has made possible the description of differential cross sections by means of the hybrid high-energy/collinear factorization within the NLO perturbative order and the NLL logarithmic accuracy. A first, systematic analysis has been performed on gauging effects rising from employing two distinct higher-order cross-section representations. One of them embodies terms which are beyond the NLL accuracy.

The study presented in this article extends our program on high-energy emissions of baryons at the LHC, started with a similar analysis on $\Lambda$ hyperons\tcite{Celiberto:2020tmb} and carried on with the discovery of the \emph{natural stability} of the high-energy resummation from $\Lambda_c$ fragmentation\tcite{Celiberto:2021dzy}.
The stabilization mechanism is connected to the behavior of the heavy-hadron gluon FF. It comes out as a general property shared by all the heavy-flavored species recently studied in the context of high-energy QCD phenomenology: single-charmed\tcite{Celiberto:2021dzy} and single-bottomed hadrons\tcite{Celiberto:2021fdp}, vector-quarkonium states\tcite{Celiberto:2022dyf,Celiberto:2022grc,Bolognino:2022paj}, and charmed $B$ mesons\tcite{Celiberto:2022keu}.
In this article a clear evidence was provided that the stabilization mechanism coming from collinear fragmentation is also present in the $\Xi$-baryon case, and has allowed us to study $\Xi$~plus~jet differential distributions around the natural energy scales provided by kinematics.

Two prospective developments are underway.
On one hand, a formal proof of the natural stability, emerged so far as a phenomenological property of semi-hard observables, needs to be provided. In particular, this will help us to shed light on the reason why the nondecreasing-with-$ \mu_F$ behavior of the gluon FF is shared also by some lighter hadron species, such as $\Xi$ particles.

On the other hand, a required step to reach the precision level in the theoretical analysis of high-energy observables relies in enhancing our hybrid factorization into a \emph{multi-lateral} and unified formalism that encodes several different resummations.
This is in line with recent studies on ultraforward emissions of light mesons\tcite{Celiberto:2022rfj} or single-charmed hadrons\tcite{Celiberto:2022zdg} at the planned Forward Physics Facility\tcite{Anchordoqui:2021ghd,Feng:2022inv}. There, it was highlighted that, although resummed distributions are stable under scale variations, they still exhibit a sensitivity when passing from a pure LL to a full NLL treatment. This is due to the simultaneous presence of both energy logarithms and large-$x$, \emph{threshold} ones. Improving our hybrid factorization by including the latters represents an urgent task to be undertaken in the short-term future.

\clearpage

\section*{Acknowledgments}

The Author thanks Maryam Soleymaninia, Hadi Hashamipour, Hamzeh Khanpour, and Hubert Spiesberger for providing him with grids of the {\tt SHKS22} fragmentation functions\tcite{Soleymaninia:2022qjf}.
The Author would like to express is gratitude to Alessandro~Papa for a critical reading of the manuscript and for useful suggestions, and to Simone~Caletti for a discussion on jet-radius and jet-angularity resummations.
This work is supported by the Atracci\'on de Talento Grant n. 2022-T1/TIC-24176 of the Comunidad Aut\'onoma de Madrid, Spain, and by the INFN/NINPHA Project, Italy.
The Author thanks the Universit\`a degli Studi di Pavia for the warm hospitality.

\bibliographystyle{apsrev}
\bibliography{references}

\end{document}